\documentclass[nature,aps,amsfonts,floatfix,superscriptaddress,longbibliography,nofootinbib,twocolumn]{revtex4-2}
\usepackage[utf8]{inputenc}
\usepackage{amsmath}    
\usepackage{mathbbol}
\usepackage{graphicx}   
\usepackage{verbatim}   
\usepackage{color}      
\usepackage{subfigure}  
\usepackage{hyperref} 

\usepackage{multirow}

\usepackage{siunitx}
\usepackage{physics}

\usepackage{siunitx}
\usepackage{physics}
\usepackage{soul}
\usepackage{enumerate}
\usepackage{empheq}
\usepackage{multirow}

\usepackage{xr}
\usepackage{xcite}
\externalcitedocument{USEthisFILE}

\renewcommand{\figurename}{{\bf Supplementary Figure}}

\def\bibsection{\section*{Supplementary References}}

\renewcommand{\figurename}{{\bf Fig.}}

\def\bibsection{\noindent {\bf REFERENCES}}

\begin{document}

\title{Spectral kissing and its dynamical consequences in the squeeze-driven Kerr oscillator}

\author{Jorge Ch\'avez-Carlos}
\affiliation{Department of Physics, University of Connecticut, Storrs, Connecticut 06269, USA}
\author{Tal\'ia L. M. Lezama}
\affiliation{Department of Physics, Yeshiva University, New York, New York 10016, USA}
\author{Rodrigo G. Corti\~nas}
\affiliation{Department of Applied Physics and Physics, Yale University, New Haven, Connecticut 06520, USA}
\author{Jayameenakshi Venkatraman}
\affiliation{Department of Applied Physics and Physics, Yale University, New Haven, Connecticut 06520, USA}
\author{Michel H. Devoret}
\affiliation{Department of Applied Physics and Physics, Yale University, New Haven, Connecticut 06520, USA}
\author{Victor S. Batista}
\affiliation{Department of Chemistry, Yale University, 
P.O. Box 208107, New Haven, Connecticut 06520-8107, USA}

\author{Francisco P\'erez-Bernal}
\affiliation{Departamento de Ciencias Integradas y Centro de Estudios Avanzados en F\'isica, Matem\'aticas y Computaci\'on, Universidad de Huelva, Huelva 21071, Spain}
\affiliation{Instituto Carlos I de F\'isica Te\'orica y Computacional, Universidad de Granada, Fuentenueva s/n, 18071 Granada, Spain}
\author{Lea F. Santos}

\affiliation{Department of Physics, University of Connecticut, Storrs, Connecticut 06269, USA}

\begin{abstract}
Transmon qubits are the predominant element in circuit-based quantum information processing, such as existing quantum computers, due to their controllability and ease of engineering implementation. But more than qubits, transmons are multilevel nonlinear oscillators that can be used to investigate fundamental physics questions. Here, they are explored as simulators of excited state quantum phase transitions (ESQPTs), which are generalizations of quantum phase transitions to excited states. We show that the spectral kissing (coalescence of pairs of energy levels) experimentally observed in the effective Hamiltonian of a driven SNAIL-transmon is an ESQPT precursor. We explore the dynamical consequences of the ESQPT, which include the exponential growth of out-of-time-ordered correlators, followed by periodic revivals, and the slow evolution of the survival probability due to localization. These signatures of ESQPT are within reach for current superconducting circuits platforms and are of interest to experiments with cold atoms and ion traps.
\end{abstract}

\maketitle

\noindent {\bf INTRODUCTION}

\noindent Recent developments in superconducting circuits have opened the pathway to explore long standing predictions of quantum physics. They have been used to study dynamical bifurcation~\cite{Dykman1990,Siddiqi2005}, to squeeze quantum fluctuations~\cite{Castellanos2008}, to prepare exotic quantum states, and to process and stabilize quantum information~\cite{Puri2017,Grimm2020}.  Here, we propose to use this platform as a quantum simulator of excited state quantum phase transitions (ESQPTs), a phenomenon that occurs in various nuclear, atomic, molecular, and condensed matter systems. The superconducting circuit considered is a driven system, whose static effective Hamiltonian describes a double-well system and thus exhibits an ESQPT. This perspective adds another layer of interest to the long history of studies on driven nonlinear oscillators~\cite{Marthaler2006,Lin2015,Marthaler2007, Peano2012,Zhang2017,DykmanBook,Dykman2018, Wang2019PRX,Venkatraman2022}, where the emergence of a double well, reached by driving the oscillator at twice its original frequency~\cite{Marthaler2006}, has been explored in studies of quantum activation~\cite{Marthaler2006,Lin2015}, quantum tunneling~\cite{Marthaler2007, Peano2012}, and the preparation of selected superpositions of quasienergy states~\cite{Zhang2017} with applications to quantum information science, such as the generation of Schr\"odinger cat states. 

A quantum phase transition (QPT) corresponds to an abrupt change in the ground state of a physical system when a control parameter reaches a critical point. It occurs in the thermodynamic limit, but scaling analyses of finite systems can signal its presence. ESQPT is a generalization of this phenomenon to excited states~\cite{Cejnar2006,Cejnar2007,Caprio2008,Cejnar2021}, which can take place independently of the presence of QPTs ~\cite{Stransky2021,Corps2022} and can be triggered by anharmonicities~\cite{Bernal2010,Khalouf2022,Gamito2022}. In an ESQPT, the separation of the states in two phases~\cite{Corps2021} occurs at a point that depends on both the value of the energy and of the control parameter. There is a vast literature on the subject, which is reviewed in~\cite{Cejnar2021}. ESQPTs are associated with enhanced decoherence~\cite{Relano2008,PerezFernandez2009},  localized eigenstates~\cite{SantosBernal2015,Bernal2016,Santos2016}, very slow~\cite{SantosBernal2015,Bernal2016,Santos2016} or accelerated~\cite{Lobez2016,Kloc2018,Pilatowsky2020} 
quantum quench dynamics, specific dynamical features at long times~\cite{Wang2019,Wang2021phase,Kloc2021}, isomerization reactions~\cite{Khalouf2019}, and the creation of Schr\"odinger cat states~\cite{Corps2022}.   

The main signature of an ESQPT is a singularity in the density of states (DOS) that moves to higher excitation energies as the control parameter increases, and may be accompanied by the closing of energy gaps between excited states. The energy where the divergence of the DOS takes place is the ESQPT critical energy. These and related features have been theoretically identified in various quantum systems with few degrees of freedom~\cite{Cejnar2006,Cejnar2007,Caprio2008,Cejnar2021,Corps2021,Stransky2021,Corps2022,Relano2008,PerezFernandez2009,Bernal2010,Wang2019,Khalouf2022,Gamito2022,Pilatowsky2020,Wang2021phase,Kloc2021,Khalouf2019,SantosBernal2015,Bernal2016,Santos2016,Lobez2016,Kloc2018,Bernal2008,Cejnar2009,Fernandez2009,Fernandez2011,Fernandez2011b,Brandes2013,Bastarrachea2014a,Bastarrachea2014b,Stransky2014,Stransky2015,Chavez2016,Chinni2021,Leyvraz2005}, and a proposal to detect
the ESQPT with spinor Bose-Einstein condensates also exists~\cite{Feldmann2021}.

Even though spectroscopic signatures of the ESQPT have been experimentally observed~\cite{Larese2011, Larese2013, KRivera2020,Dietz2013,Zhao2014} and its presence suggested from the bifurcation phenomenon detected in~\cite{Zibold2010,Araujo2013, Trenkwalder2016}, presently none of these systems provides the means to analyze the spectrum as a function of the control parameter and to simultaneously observe the dynamical consequences of an ESQPT in a controllable way. Superconducting circuits close this gap by offering a platform that has an experimental realizable classical limit and provides both frequency- and time-resolved high quantum non-demolition measurements fidelity~\cite{Frattini2022}.

As we explain here, the exponential approach of pairs of adjacent levels (spectral kissing) recently observed in the spectrum of the superconducting Kerr resonator as a function of the amplitude of a squeezing drive~\cite{Frattini2022}, and previously discussed in~\cite{Zhang2017}, marks the presence of an ESQPT. The dynamical counterpart of this transition presents a seeming paradoxical behavior, which can, in principle, be observed in a system such as the one in Ref.~\cite{Frattini2022}. For Glauber coherent states close to the ESQPT, the initial decay of the survival probability (overlap of the initial and the evolved state) is slower than for coherent states away from the ESQPT, while the fidelity out-of-time-ordered correlator (FOTOC) grows exponentially fast for the first and slower for the latter. The justification for these apparently opposite behaviors lies in the classical limit of the system. At the origin of the phase space, $(q=0,p=0)$, there is a stationary but unstable point that is associated with the ESQPT. At this point, the evolution is dominated by the squeezing part of the Hamiltonian. 

The experimental capability of reconstruction of the full phase-space distribution~\cite{Frattini2022} motivates our analysis of the dynamics in phase space, which reveals features that were missed by previous works on ESQPTs and that are of interest to studies of nonequilibrium quantum dynamics. Depending on the initial state, the exponentially fast spread in phase space can be followed by the onset of complicated interference patterns or yet by periodic revivals that persist for long times. Our analysis also elucidates why states with exactly the same energy may exhibit different dynamics.

\vskip 0.4 cm 
\noindent {\bf RESULTS}

\noindent {\bf Quantum system.} The system that we investigate was implemented in a superconducting circuit~\cite{Frattini2022} based on driven SNAIL~\cite{Frattini2017} transmons. The static effective Hamiltonian of this system is given by (Supplementary Note 1)
\begin{equation}
\frac{\hat{H}_{qu}}{\hbar\, K} =   \hat{n} (\hat{n}-1) -  \xi \left( \hat{a}^{\dagger 2} +  \hat{a}^2  \right),
\label{Eq:Ham}
\end{equation}
where $\hat{n}=\hat{a}^{\dagger} \hat{a}$, $K$ is the Kerr nonlinearity, $\xi = \epsilon_2/K$ is the control parameter, and $\epsilon_2$ is the squeezing amplitude. The system conserves parity, $[\hat{H}_{qu},(-1)^{\hat{a}^{\dagger} \hat{a}}]=0$.

We study the spectrum of $\hat{H}_{qu}$  as a function of the control parameter $\xi$ in Figs.~\ref{fig01}a-e. The plots display the excitation energies, $E'=(E-E_0)$, where $E$ are the eigenvalues of $\hat{H}_{qu}$ and $E_0$ its  ground state energy. The numerical data in Fig.~\ref{fig01}a  reproduce the experimental data in Fig.~3A of Ref.~\cite{Frattini2022}. One sees that as the control parameter increases, the coalescence of a pair of adjacent eigenvalues,  each level belonging to a different parity sector, happens at a higher energy. This spectral kissing becomes better visible in Fig.~\ref{fig01}b, where larger values of $\xi$ are used. For a given value of the control parameter, the spectral kissing happens at the critical energy of the ESQPT, $E'_{\text{ESQPT}}$, which is marked with a solid line in Fig.~\ref{fig01}b and is obtained analytically [see Eq.~(\ref{Eq:E_ESQPT}) below]. 
 
In addition to the exponential approach of the energies in each pair, the eigenvalues cluster at $E'_{\text{ESQPT}}$ ( Supplementary Note 2). This produces the peak of the DOS displayed for different values of the control parameter in Figs.~\ref{fig01}c-e. The peak diverges for $\xi \rightarrow \infty$, which is a main signature of the ESQPT~\cite{Caprio2008}. 


	\begin{figure*}[t!]
 \bf{
	\centering
	\includegraphics[width=0.65\textwidth]{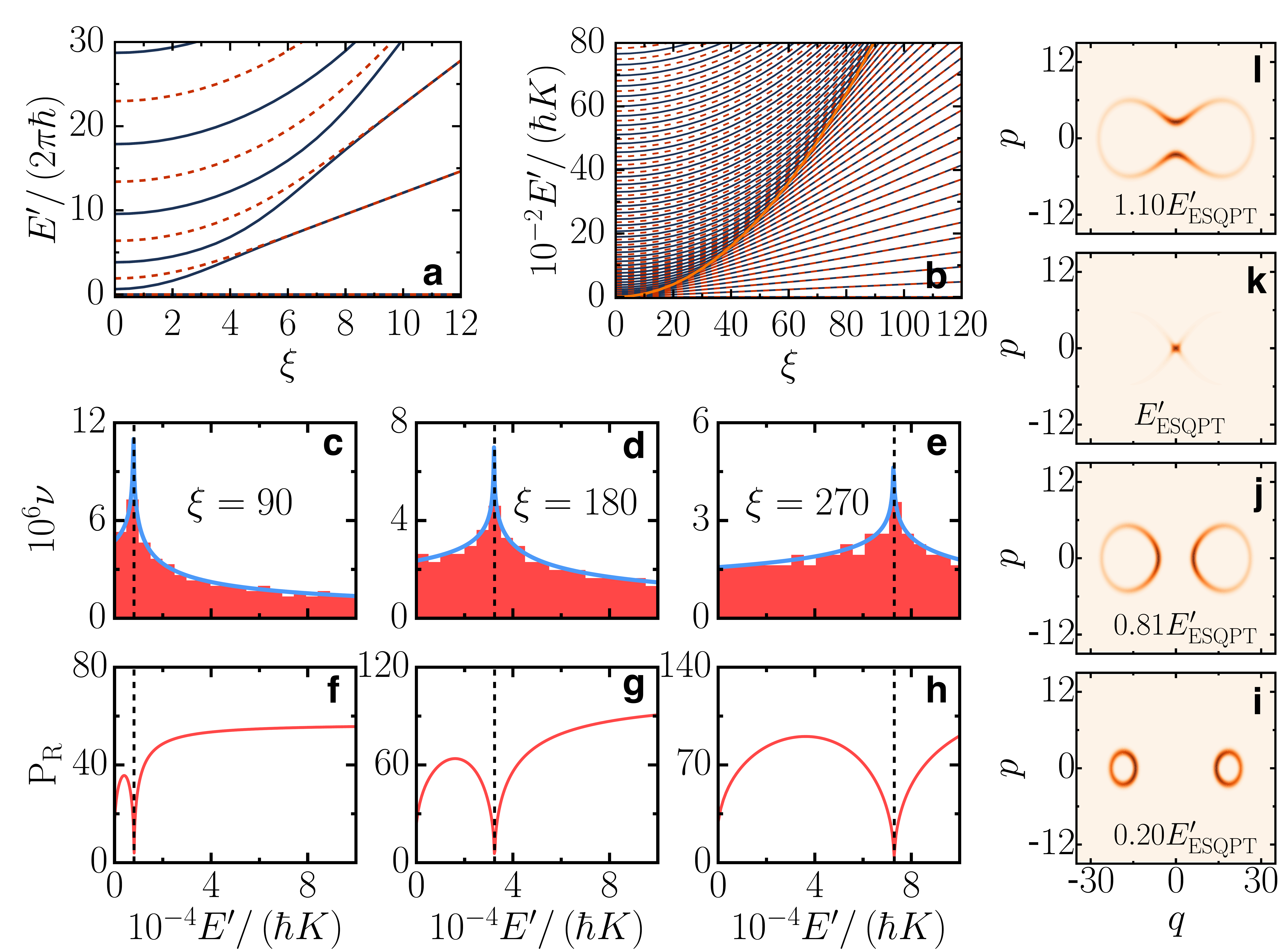}}
	\caption{{\bf Spectral kissing and localization.} {\bf a} {\normalfont Energy levels as a function of the control parameter reproducing the experimental data~\cite{Frattini2022} with $K/(2\pi)=0.32$MHz and {\bf b} $E'/(\hbar K)$ for larger values of $\xi$. Solid lines are for the even parity sector and dashed lines for odd parity. The bright orange line in {\bf b} marks the energy of the ESQPT, as given in Eq.~(\ref{Eq:E_ESQPT}). {\bf c-e} Normalized density of states and {\bf f-h} participation ratio for the eigenstates in the Fock basis for the values of $\xi$ indicated in {\bf c-e};  even parity sector. Numerical (shade) and analytical (solid line) data are shown in {\bf c-e}. The vertical dashed line in} {\bf c-h} {\normalfont is the ESQPT energy from Eq.~(\ref{Eq:E_ESQPT}). {\bf i-l} Husimi functions for different eigenstates and  $ \xi =180$.}
	}
		\label{fig01}
	\end{figure*}

The presence of the ESQPT gets reflected in the structure of the eigenstates, $|\psi \rangle = \sum_n C_n |n \rangle$, written in the Fock basis, $\hat{a}^{\dagger} \hat{a}|n \rangle = n |n \rangle$. The eigenstates at the vicinity of the ESQPT are highly localized in the Fock state $|0\rangle$ \cite{SantosBernal2015, Santos2016, Bernal2016}. This can be quantified with the participation ratio,
$\text{P}_{\text{R}} =1/\sum_{n=0}^{{\cal N}-1} | C_{n}|^4 $,
where ${\cal N}$ is the size of the truncated Hilbert space. $\text{P}_{\text{R}}$ is large for an extended state and small for a localized state. In Figs.~\ref{fig01}f-h, we show the participation ratio as a function of $E'$. An abrupt dip in the value of $\text{P}_{\text{R}}$ happens for $E' \sim E_{\text{ESQPT}}'$ and the analysis of the components of the eigenstate at this energy confirms its localization at $|0\rangle$. Equivalently to $\text{P}_{\text{R}}$, the plot of the occupation number $\langle \psi |\hat{a}^{\dagger} \hat{a}|\psi \rangle$ as a function of energy exhibits a dip at $ E' \sim  E_{\text{ESQPT}}'$ (Supplementary Note 2).

The localization at the ESQPT critical point is also detected with the Husimi function~\cite{Wang2021} obtained by writing the eigenstates in the basis of Glauber coherent states [see Eq.~(13)]. The Husimi function gives the distribution of the quantum state in the phase space of canonical variables $(q,p)$. As seen in Fig.~\ref{fig01}k, the eigenstate closest to the ESQPT energy is highly concentrated in the origin of the phase space. This contrasts with the eigenstates below the ESQPT [Figs.~\ref{fig01}i-j], which present two separated ellipses, and the eigenstates above it [Fig.~\ref{fig01}l]. The localization in the phase space mirrors the localization in the Fock basis, since the coherent state with $(q=0,p=0)$ coincides with the Fock state $|0\rangle$.

\vskip 0.4 cm  

\noindent {\bf Classical limit.} The Hamiltonian of the Kerr oscillator in Eq.~(\ref{Eq:Ham}) develops two wells when $\xi>0$. The depth of the wells and their energy levels grow as $\xi$ increases, bringing the system closer to the classical limit. Experimentally, the value of $\xi$ can be increased by reducing the impedance of the circuit, increasing the microwave power of the squeezing drive, or approaching the Kerr-free point (Supplementary Note 1). 

The grounds for the onset of the ESQPT are found in the classical limit. The classical Hamiltonian is derived in Methods and is given by
\begin{eqnarray}
\frac{H_{cl}}{K}  =  \frac{1}{4} (q^2 +  p^2 )^2   -  \xi (q^2  - p^2) .
\label{EqHcl}
\end{eqnarray} 
It presents three stationary points when $\xi>0$. They are the two center points $\{ q, p \}= \{ \pm \sqrt{2  \xi} , 0\} $   with the minimal energy of the system $ {\cal E}_{\text{min}} = H_{cl} (q,p)= - K  \xi^2$, and the hyperbolic point $\{ q, p \}= \{ 0 , 0\} $ with energy  ${\cal E}_{\text{hyp}} = 0$. In the plot of the energy contours in Fig.~\ref{fig02}a, the hyperbolic point is indicated as O, the red line that intersects at this point is the separatrix, and the two blue diamonds are the center points. 

The properties of the quantum system find a parallel in the classical limit. The energy difference ${\cal E}_{\text{hyp}} - {\cal E}_{\text{min}}$ marks the separatrix in Fig.~\ref{fig02}a and determines the energy of the ESQPT, 
\begin{equation}
    E_{\text{ESQPT}}' \approx  K\xi^2 ,
     \label{Eq:E_ESQPT}
\end{equation}
which is indicated with a bright orange line in Fig.~\ref{fig01}b. The equality in Eq.~(\ref{Eq:E_ESQPT})  holds in the classical limit. Below this energy, the pairs of stable periodic orbits with equal energy are analogous to the degenerate states of the quantum system, and above that line the degeneracy is lost. The stationary point at the origin of the phase space,  $(q,p)=(0,0)$,  justifies the localization at the Fock state $|0\rangle$ of the eigenstate with energy at the ESQPT.

The existence of a non-degenerate hyperbolic point implies the logarithmic discontinuity of the level density, as shown in Refs.~\cite{Stransky2016,Cejnar2021}, and explains the peak at $E'_{\text{ESQPT}}$ in Figs.~\ref{fig01}c-e. Using the smooth component of the Gutzwiller trace formula~\cite{GutzwillerBook}, we obtain a semiclassical approximation for the DOS (Supplementary Note 3). This curve outlines the numerical data in Figs.~\ref{fig01}c-e. 

Another consequence of the hyperbolic point is the onset of a positive Lyapunov exponent (Supplementary Note 4),
\begin{equation}
\lambda = 2 K \xi .
\label{Eq:lambda}
\end{equation}
The system described by Eq.~(\ref{EqHcl}) is regular, so the Lyapunov exponent for any initial condition is zero, except for the unstable point O \cite{Hummel2019,Pilatowsky2020,Kidd2021}. 

\begin{figure*}[t]
\bf{
\centering
    \includegraphics[width=0.8\textwidth]{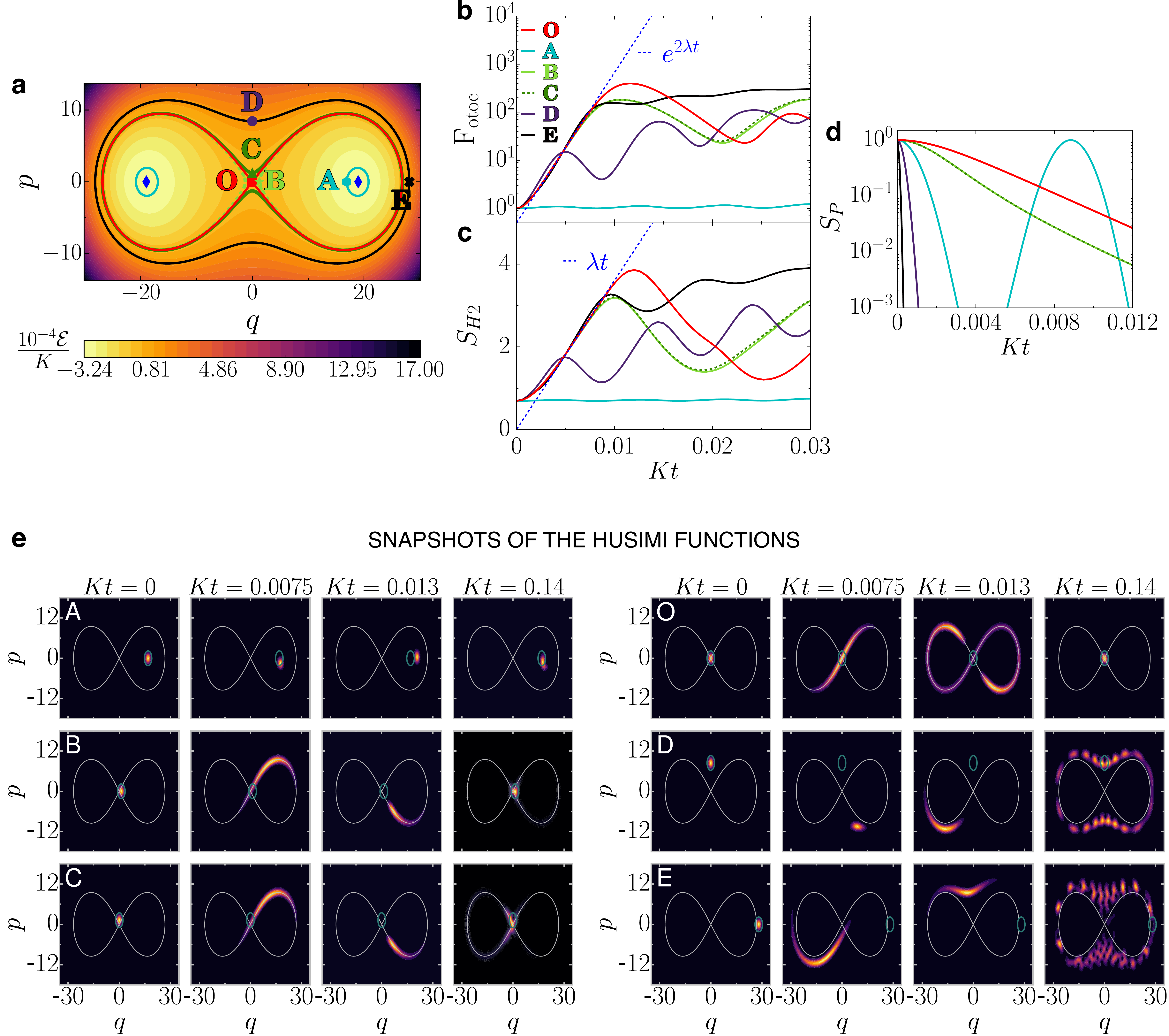}}
	\caption{{\bf Phase space and quantum dynamics.} {\normalfont{\bf a} Energy curves in the phase space obtained with Eq.~(\ref{EqHcl}). The hyperbolic point is denoted as O, the center points are represented with blue diamonds, and the solid line intersecting at O is the separatrix. Points O, A-E mark the centers of the initial coherent states chosen for the quantum dynamics.  {\bf b} Evolution of the FOTOC,  {\bf c} Husimi entropy, and  {\bf d} survival probability as a function of time. The exponential [linear] curve with rate [slope] given by the Lyapunov exponent in Eq.~(\ref{Eq:lambda}) are indicated in {\bf b} [{\bf c}]. {\bf e} Snapshots of the Husimi functions; each row refers to one of the six initial coherent states investigated, and each column to a different time, as indicated.}
	}
		\label{fig02}
	\end{figure*}

\vskip 0.4 cm  

\noindent {\bf Quantum dynamics: Instability.} The instability associated with the hyperbolic point is manifested in the quantum domain with the exponential growth of out-of-time-ordered correlators (OTOCs) \cite{Hummel2019,Pilatowsky2020,Hashimoto2020,Kidd2021}. These quantities, defined as $O_{\text{toc}}=\langle [\hat{W}(t), \hat{V}(0) ]^2\rangle$, measure the spread (scrambling) of quantum information by assessing how the operators $\hat{W}$ and $\hat{V}$ fail to commute due to the evolution of $\hat{W}$ \cite{Maldacena2016JHEP}. A particular example of OTOCs is the FOTOC, which corresponds to having the operator $\hat{V}=|\Psi(0)\rangle \langle \Psi(0)|$, for the initial state $|\Psi(0)\rangle $, and $\hat{W}=\text{e}^{i \delta \phi \hat{G}}$, where $\delta \phi$ is a small perturbation and $\hat{G}$ is a Hermitian operator. In the perturbative limit, $\delta \phi \ll 0$, the FOTOC is the variance $\sigma_G^2(t) = \langle \hat{G}^2(t) \rangle - \langle \hat{G}(t) \rangle^2$ \cite{Lewis-Swan2019}.

We analyze the evolution of the FOTOC given by the variance of $p$ and $q$, 
\begin{equation}
    F_{\text{otoc}} (t) =\sigma_p^2(t) + \sigma_q^2(t), 
    \label{Eq:FOTOC}
\end{equation}
because the initial coherent states that we consider spread in both canonical coordinates~\cite{Pilatowsky2020}. These states are centered at the points O, A-E, marked in Fig.~\ref{fig02}a,  and are denoted as $|\Psi_j(0)\rangle$ with $j = O, A, \ldots, E$.   State $|\Psi_A (0)\rangle$ has the lowest energy,  followed by $|\Psi_B (0)\rangle$ (negative energy close to zero), $|\Psi_O (0)\rangle$ (zero energy), and $|\Psi_C (0)\rangle$ (positive energy close to zero). States $|\Psi_D (0)\rangle$ and $|\Psi_E (0)\rangle$ have equal and high positive energy (see  Methods). 

We compare the growth of $F_{\text{otoc}} (t)$ in Fig.~\ref{fig02}b with the
Husimi entropy,
\begin{equation}
S_{\text{H2}} (t)=-\ln M_2(t),
\end{equation} 
 in Fig.~\ref{fig02}c, 
where $M_2(t)$ is the integral of the square of the Husimi function (Supplementary Note 5.1). Both quantities, $F_{\text{otoc}} (t)$ and $S_{\text{H2}} (t)$, measure how an evolving state spreads in the phase space. Snapshots of the evolution of the Husimi functions for $|\Psi_{A,B,C} (0)\rangle$ (left) and for $|\Psi_{O,D,E} (0)\rangle$ (right) are presented in Fig.~\ref{fig02}e (more snapshots are in Supplementary Note 5.1 and videos are available in~\cite{KerrAnimations}). The results are as follows.

(O): After the parabolic increase in $t$, that happens for very short times $Kt< K\tau =(\sqrt{8}\xi)^{-1}$ (Supplementary Note 6), $F_{\text{otoc}}^{(O)} (t)$ [$S_{\text{H2}}^{(O)}(t)$] for the initial coherent state at the hyperbolic point, $ |\Psi_{O} (0)\rangle$, grows exponentially [linearly] fast with a rate proportional to the classical Lyapunov exponent given in Eq.~(\ref{Eq:lambda}), that is,  $F_{\text{otoc}}^{(O)} (t) \propto \text{e}^{2 \lambda t}$ [$S_{\text{H2}}^{(O)} (t) \propto \lambda t$]. The snapshot of the Husimi function for a time as small as $Kt = 0.013$  indicates that $ |\Psi_{O} (t)\rangle$ is already very spread out in phase space, covering an area larger than that for the other five states, even those with larger energies. Indeed, around $Kt = 0.013$,  $F_{\text{otoc}}^{(O)} (t)$  [$S_{\text{H2}}^{(O)}(t)$] reaches the highest value among the states considered, as seen in  Fig.~\ref{fig02}b[c]. The maximum value happens at the  Ehrenfest time, $ {\cal T} \sim \ln (\xi)/\lambda$  (Supplementary Note 7). 

The fast scrambling of quantum information for $ |\Psi_{O} (t)\rangle$, which happens for $\tau<t<{\cal T}$, is later followed by partial reconstructions of the initial distribution (see the Husimi function  at $Kt = 0.14$). In the absence of dissipation, this yo-yo process of spreading and contraction persists for a long time  (Supplementary Note 5.2). This behavior is the quantum counterpart of the classical dynamics at the vicinity of the hyperbolic (saddle) point O, which is both a repellor and an attractor (Supplementary Note 4), resulting in trajectories that move both towards and away from O. We also note that despite reaching the highest value at $t \sim  {\cal T}$,  the infinite-time average of  $F_{\text{otoc}}^{(O)} (t)$ is actually smaller than the saturation value for $F_{\text{otoc}}^{(D,E)} (t)$ (Supplementary Note 7). This result shows that the degree of spreading quantified by OTOCs depends not only on the initial state and system, but also on the timescale.

(A): The initial coherent state $ |\Psi_A (0)\rangle$ is very close to a center point, so the evolution is very slow,  $F_{\text{otoc}}^{(A)} (t)$ [$S_{\text{H2}}^{(A)}(t)$] never reaches large value, and the Husimi function remains close to the point A.

(B) \& (C): State $|\Psi_{B}(0)\rangle$ [$|\Psi_{C}(0)\rangle$] is slightly below [above] the ESQPT. Instead of the confinement around the center point imposed to the  classical orbit B, quantum effects allow $|\Psi_{B}(t)\rangle$ to escape and evolve similarly to $|\Psi_{C}(t)\rangle$. The spread of the Husimi distributions for both states is comparable, reaching regions of the phase space with $+q$ and $-q$ (see snapshots in Fig.~\ref{fig02}e and in Supplementary Note 5.2). In addition, since B and C are in the vicinity of the unstable point O,  quantum fluctuations trigger the exponential [linear] growth of $F_{\text{otoc}}^{(B,C)} (t)$  [$S_{\text{H2}}^{(B,C)} (t)$] observed in  Fig.~\ref{fig02}b [Fig.~\ref{fig02}c]. This behavior is at odds with the classical limit, where the positive Lyapunov exponent emerges only at the hyperbolic point and not close to it. As $\xi$ increases and one approaches the classical limit, the duration of the exponential behavior for $F_{\text{otoc}}^{(B,C)} (t)$ decreases. 

(D) \& (E): States $|\Psi_{D} (0)\rangle$ and $|\Psi_{E}(0)\rangle$ have the same high energy, but evolve differently. In terms of scrambling, $|\Psi_{E}(0)\rangle$ combines the best of both worlds, because in addition to high energy, which leads to the largest saturation value for $F_{\text{otoc}}^{(D,E)}(t)$  (Supplementary Note 7), it partially overlaps with the separatrix (see the snapshot of the Husimi function at $t=0$ in Fig.~\ref{fig02}e), so $F_{\text{otoc}}^{(E)}(t)$ [$S_{\text{H2}}^{(E)}(t)$] in Fig.~\ref{fig02}b [Fig.~\ref{fig02}c] presents an exponential [linear] growth  analogous to that seen for $|\Psi_{B,C}(0)\rangle$, which is absent for $|\Psi_{D} (0)\rangle$.
The spread of the Husimi distribution for $|\Psi_{E}(0)\rangle$ happens simultaneously inside and outside the separatrix (Supplementary Note 5.2), leading to complicated quantum interference effects, as those observed in the snapshot of the Husimi function  at $K t =0.14$.

\vskip 0.4 cm  

\noindent {\bf Quantum dynamics: Localization.} While the fastest and longest scrambling happens for the initial coherent state $|\Psi_{O} (0) \rangle$, this state also presents the slowest decay of the survival probability,
\begin{equation}
    S_p(t) = \left| \langle \Psi(0)|\Psi(t) \rangle \right|^2.
\end{equation}
The survival probability for all other initial coherent states, with energy above or below the ESQPT, decays faster than $S_p^{(O)}(t)$, as seen in Fig.~\ref{fig02}d.

The apparent paradox of the fast spread of $|\Psi_{O} (t)\rangle$, as measured by $F_{\text{otoc}}^{(O)} (t)$ and $S_{\text{H2}}^{(O)} (t)$,  and the slow decay of $S_p^{(O)} (t)$ is naturally resolved in view of the classical limit and from the analysis of the Husimi functions. The instability associated with the hyperbolic point O is the source of the exponentially fast spread of the variance of the phase-space distribution, 
but O is also a stationary point (the gradient of the Hamiltonian at this point is zero), so $|\Psi_{O} (0)\rangle$ is strongly localized in the eigenstate at the ESQPT [see Fig.~\ref{fig01}k]. In other words, the width of the energy distribution for $|\Psi_{O} (0)\rangle$, given by $\sqrt{2}K\xi$, is the smallest one among the six states (Supplementary Note 6). Close to the origin of the phase space, the evolution is dominated by the squeezing, $\hat{H}_{qu} \approx \epsilon_2  (\hat{q}^2 - \hat{p}^2)$. This leads to the rapid stretching of $|\Psi_{O} (t) \rangle$, while part of the population remains for some time in the vicinity of the origin.
These two aspects of the dynamics become evident in the snapshot of the Husimi function for $|\Psi_{O} (t) \rangle$ at $Kt =0.0075$. The small green ellipse in those panels indicates the size of the initial coherent state. One sees that the Husimi distribution for $|\Psi_{O} (t)\rangle$ at $Kt =0.0075$ is stretched out, but part of it remains inside the green ellipse.

\vskip 0.4 cm  

\noindent {\bf DISCUSSION} 

\noindent This work bridges communities working on superconducting circuits, ESQPTs, and nonequilibrium quantum dynamics. The squeeze-driven Kerr oscillator is an addition to the list of nuclear, molecular, and condensed matter systems that exhibit ESQPTs. Its advantage is to be experimentally realizable in an available superconducting circuit platform, where both frequency and time domain measurements can be done simultaneously, the control parameter can be tuned to approach the classical limit, arbitrary initial states can be prepared, and the dynamics can be studied in phase space. We expect superconducting circuits to become versatile quantum simulators for ESQPTs and related phenomena, such as isomerization, where the separation between neighboring energy levels decreases close to the isomerization barrier height~\cite{Baraban2015,Videla2018}.

The dynamical consequences of ESQPTs that we presented should also appeal to experimental platforms, where long-range couplings can be tuned to approach models with collective interactions, such as those with cold atoms~\cite{Li2023} and trapped ions~\cite{Smith2015}. Of interest to those experiments is the demonstration of the exponential growth of OTOCs, which we showed to emerge for different initial states placed close to the separatrix that marks the ESQPT. Other highlights include the later revivals of a coherent state initially centered at the phase-space origin, the combined effects of fast scrambling and subsequent interferences for a high-energy state close to the separatrix, and the different dynamics for states with the same energy but initially located in different regions of the phase space.

We conclude with a brief discussion about the static effective Hamiltonian, $\hat{H}_{qu}$,  investigated here and used to describe the driven SNAIL transmon  in~\cite{Frattini2022}. As the drive amplitude and nonlinearities of the experimental system increase, $\hat{H}_{qu}$ ceases to be valid, the ESQPT melts away, and chaos eventually sets in. The emergence of chaos, which could be captured experimentally and may affect the development of quantum devices, cannot be described by any static effective Hamiltonian~\cite{Peano2012,Venkatraman2022,Frattini2022} obtained for systems with only one degree of freedom. The analysis of chaos, which will be the subject of our forthcoming papers, has to rely entirely on the original time-dependent Hamiltonian.

\vskip 0.4 cm  
\noindent {\bf METHODS} 

In the Supplementary Note 1, we describe how the original driven Hamiltonian leads to the static effective Hamiltonian,
\begin{equation}
\label{eq:HSK}
 \frac{\hat{H}_{qu}}{\hbar} =  - K \hat{a}^{\dagger 2} \hat{a}^2 + \epsilon_2 (\hat{a}^{\dagger 2} +  \hat{a}^2),    
\end{equation}
and how the parameters can be experimentally controlled. In the main text, we changed the sign of the Hamiltonian in Eq.~(\ref{Eq:Ham}) for convenience, so that we could say that  $E_0$ in $E'=E-E_0$ is the ground state energy of $\hat{H}_{qu}$, instead of its highest energy. Regardless of the sign convention, dissipation will bring the experimental system to the attractors (stable nodes) in the bottom of the wells, which define unambiguously the ground state of the system.

\vskip 0.3 cm 
\noindent {\bf Classical limit}

\noindent For large values of the control parameter, $\xi=\epsilon_2/K\gg1$, the double wells created by the quantum Hamiltonian in Eq.~(\ref{eq:HSK}) become very deep and the number of levels inside the wells become macroscopic, so $\hat{H}_{qu}$ exhibits properties comparable to the classical Hamiltonian. However, to derive the classical Hamiltonian for any depth of the wells, that is, to approach a continuous spectrum for a fixed and not necessarily large value of the control parameter,  we introduce the parameter $N_{\text{eff}}$, whose reciprocal is related with the size of the zero point fluctuations. We write
\begin{equation}
\label{eq:a_z}
\hat{a}= \sqrt{ \frac{N_{\text{eff}}}{2 } }\left(\hat{q}+i  \hat{p}\right),
\end{equation}
and
$$
[\hat{q},\hat{p}]=\frac{i}{N_{\text{eff}} } ,
$$
so the classical limit can be reached by taking $N_{\text{eff}} \rightarrow \infty$, since $\hat q\rightarrow q$ and $\hat p\rightarrow p$. This way, the quantum Hamiltonian,
\begin{eqnarray}
\frac{H_{qu}}{\hbar } &=&  -  \frac{K N_{\text{eff}}^2}{4 } \left(\hat{q}-i  \hat{p}\right)^2 \left(\hat{q}+i  \hat{p}\right)^2 \nonumber \\
 &+& \xi \frac{K N_{\text{eff}}}{2 }[ \left(\hat{q}-i  \hat{p}\right)^2 +\left(\hat{q}+i  \hat{p}\right)^2],
\end{eqnarray}
leads to the classical Hamiltonian (with $\hbar = 1$),
\begin{equation}
 H_{cl} =  -  \frac{K_{cl}}{4 } (q^2+p^2)^2 + K_{cl} \xi_{cl} ( q^2-p^2) ,
 \label{EqHcl_SM}
\end{equation}
where 
$$K=K_{cl}/N_{\text{eff}}^2 \hspace{0.3 cm} \text{and} \hspace{0.3 cm} \xi = \xi_{cl} N_{\text{eff}}.$$ In the main text, we fixed 
$$N_{\text{eff}}=1,$$ 
and used large values of $\xi$. 

The experimental system admits an approximate classical description if it is initialized in a coherent state and for as long as the Hamiltonian phase space surface produces only a linear force (a quadratic Hamiltonian) over the spread of the evolving state.

\vskip 0.3 cm 
\noindent {\bf Husimi Function}

\noindent For an eigenstate written in the basis of the Glauber coherent states, 
\begin{equation}
|\alpha\rangle = \text{e}^{-\frac{1}{2} |\alpha|^2} \sum_{n=0}^{{\cal N}} \frac{\alpha^n}{\sqrt{n!}} |n \rangle ,
\label{EqSM_Coh}
\end{equation}
where  $\hat{a} |\alpha\rangle = \alpha|\alpha\rangle$, ${\cal N}$ is the truncation of the Hilbert space,  
$$\alpha = \sqrt{\frac{1}{2} } (q+i p) $$ 
and $N_{\text{eff}}=1$, the Husimi function is given by
\begin{equation}
Q^{\psi}(q,p)  \!=\! \frac{1}{2 \pi } \left| \sum_{n=0}^{\cal N} C_{n} \text{e}^{-  \frac{(q^2+p^2)}{4}  }\frac{(q-ip)^n }{\sqrt{2^n  n!}} \right|^2 .
\label{EqSM_Q}
\end{equation}  

\vskip 0.3 cm 
\noindent {\bf Initial Coherent States}

\noindent
The six initial coherent states that we consider are obtained by using in Eq.~(\ref{EqSM_Coh}) the values of $p$ and $q$ specified below. These are the points marked in Fig.~2a. Their classical energies ${\cal E}$ are given for $\xi_{cl}=180$.

\begin{eqnarray}
\text{Point O}:  && \hspace{0.3 cm} q=0, \, p=0, \nonumber \\ 
&& \hspace{0.3 cm} {\cal E}/K_{cl}=0. \nonumber \\
&& \nonumber 
\\
\text{Point A}: && \hspace{0.3 cm} q=16.9143, \, p=0, \nonumber \\
&& \hspace{0.3 cm} {\cal E}/K_{cl}=-3.1034 \times10^4 . \nonumber \\
&& \nonumber
\\
\text{Point B}: && \hspace{0.3 cm} q=1.2533, \, p=0, \nonumber \\
&& \hspace{0.3 cm} {\cal E}/K_{cl}=-0.0282 \times10^4 . \nonumber \\
&& \nonumber
\\
\text{Point C}: && \hspace{0.3 cm} q=1.2506, \, p=0, \nonumber \\
&& \hspace{0.3 cm} {\cal E}/K_{cl}=0.0282 \times10^4 . \nonumber \\
&& \nonumber
\\
\text{Point D}: && \hspace{0.3 cm} q=0, \, p=8.4443, \nonumber \\
&& \hspace{0.3 cm} {\cal E}/K_{cl}=1.4106 \times10^4 . \nonumber \\
&& \nonumber
\\
\text{Point E}: && \hspace{0.3 cm} q=28.1302, \, p=0, \nonumber \\
&& \hspace{0.3 cm} {\cal E}/K_{cl}=1.4106 \times10^4 .  
\end{eqnarray}


\vskip 0.3 cm 
\noindent {\bf DATA AVAILABILITY}

\noindent
All data for Fig.1 and Fig.2 can be downloaded from 
\url{https://www.dropbox.com/scl/fi/0tggwm9wyjiknrwmx1o8x/DATA_npjQuantInf.zip?rlkey=4stxzad21bmk7fijh79yiwc6a&dl=0} 
or from 
\url{https://gitlab.com/currix1/kerr\_resonator\_animations}. 

\vskip 0.3 cm 
\noindent {\bf CODE AVAILABILITY}

\noindent
All the computational codes that were used to generate the
data presented in this paper are available from the corresponding authors upon request.

\vskip 0.4 cm 


%

\vskip 0.3 cm 
\noindent  {\bf ACKNOWLEDGMENTS}

\noindent
This research was supported by the NSF CCI grant (Award Number 2124511). T.L.M.L. was funded by the NSF grant No. DMR-1936006. F.P.B. was funded by the I+D+i project PID2019-104002GB-C21 (MCIN/AEI/10.13039/501100-\\
\noindent
011033) and by the Consejer\'{\i}a de Conocimiento, Investigaci\'on
y Universidad, Junta de Andaluc\'{\i}a and European Regional
Development Fund (ERDF), ref.~UHU-1262561. Computing resources supporting
this work were partly provided by the CEAFMC and Universidad de Huelva High
Performance Computer (HPC@UHU) located in the Campus Universitario el
Carmen and funded by FEDER/MINECO project UNHU-15CE-2848. L.F.S. had support from the MPS Simons Foundation Award ID: 678586. J.C.C. and L.F.S. thank Jorge Hirsch and his group for various discussions on OTOCs and ESQPTs.

\vskip 0.3 cm 
\noindent {\bf AUTHOR CONTRIBUTIONS}

\noindent J.C.-C., F.P-B., and L.F.S. conceived the original ideas. J.C.-C., F.P-B., and L.F.S. drafted the manuscript and shaped it with help from T.L.M.L and R.G.C. The calculations were performed by J.C.-C. and F.P-B. with support from T.L.M.L. and L.F.S. The discussions had participation of all authors, including J.V., M.H.D., and V.S.B., who gave useful suggestions. 

\vskip 0.3 cm 
\noindent  {\bf COMPETING INTERESTS}

\noindent
The authors declare no competing interests.

\vskip 0.3 cm 
\noindent {\bf ADDITIONAL INFORMATION}

\noindent {\bf Supplementary information} is available for this paper.

\noindent {\bf Correspondence} and requests for materials should be addressed to Lea F. Santos.

\vskip 5 cm 
\onecolumngrid

\begin{center}

{\large \bf Supplementary Information: Spectral kissing and its dynamical consequences in the squeeze-driven Kerr oscillator}

\vspace{0.5cm}
Jorge Ch\'avez-Carlos$^{1}$, Tal\'ia L.M. Lezama$^{2}$, Rodrigo G. Corti\~nas$^{3}$, Jayameenakshi Venkatraman$^{3}$, Michel H. Devoret$^{3}$, Victor S. Batista$^{4}$, Francisco P\'erez-Bernal$^{5,6}$ and Lea F. Santos$^1$\\ 
\vskip 0.1cm 
$^1${\it Department of Physics, University of Connecticut, Storrs, Connecticut, USA}\\
$^2${\it Department of Physics, Yeshiva University, New York, New York 10016, USA}\\
$^3${\it Department of Applied Physics and Physics, Yale University, New Haven, Connecticut 06520, USA}\\
$^4$ {\it Department of Chemistry, Yale University, P.O. Box 208107, New Haven, Connecticut 06520-8107, USA}\\
$^5$ {\it Departamento de Ciencias Integradas y Centro de Estudios Avanzados en F\'isica,
Matem\'aticas y Computaci\'on, Universidad de Huelva, Huelva 21071, Spain}\\
$^6$ {\it Instituto Carlos I de F\'isica Te\'orica y Computacional,
Universidad de Granada, Fuentenueva s/n, 18071 Granada, Spain}
\end{center}

\maketitle
This Supplementary Information is organized as follows. Supplementary Note 1 contains details about the quantum and classical Hamiltonians of the squeeze-driven Kerr oscillator and a discussion about the experimental parameters. Supplementary Note 2 compares two plots that present the excitation energies $E'$ as a function of the control parameter $\xi$. In one plot both parities are included and in the other one, only one parity sector is considered. In addition, this Supplementary Note includes a figure for the occupation number, which also detects the ESQPT. Supplementary Notes~3 and 4 give the density of states (DOS) and the Lyapunov exponent, respectively. Supplementary Note 5 provides an equation for the integral of the square of the Husimi function and additional snapshots for the evolution of the Husimi functions for the six initial coherent states studied in the main text. In Supplementary Note 6, we discuss how to derive the time interval for the initial quadratic behavior in $t$ of the survival probability, FOTOC, and $M_2(t)$. Supplementary Note 7 shows the duration of the exponential growth of the FOTOC for the coherent state O and the saturation values of the FOTOC for the six initial coherent states that we study.

\setcounter{figure}{0}

\section{Quantum and Classical Hamiltonians}
\label{SM_ham}

In the same way that an LC circuit is the electrical analog of a mechanical harmonic oscillator, the Josephson junction is the electrical analog of a mechanical pendulum. The Hamiltonian of a single Josephson junction is \cite{devoret1995quantum,Blais2022}
$$\hat H  = \frac{1}{2C}\hat Q^2 - E_J\cos \left( \frac{2\pi}{\Phi_0} \hat{\Phi} \right),$$
where $C$ is the circuit's capacitance, $E_J$ is the Josephson energy, $\hat{\Phi}$ is the phase circuit variable, and $\hat{Q}$ its charge, with  $[\hat{\Phi},\hat{Q}]=i\hbar$  \cite{Blais2022}. This is the canonical commutation relation that describes quantum circuits and is analogous to the position-momentum relation in a mechanical system. The charge enters the Hamiltonian as a quadratic kinetic energy and the circuit's phase enters via the Josephson cosine potential and is analogous to the projection of a constant gravitational field over the vertical as in a pendulum potential \cite{Girvin2014}.  

One defines the bosonic operators of the circuit as a convenient calculation tool. The annihilation operator for a superconducting circuit takes the form 
\begin{equation}
\label{eq:a_z}
\hat{a}=\sqrt{\frac{1}{ 2 \hbar Z}}\left(\hat{\Phi}+i Z \hat{Q}\right)
\end{equation}
where $Z$ is the impedance of the circuit  and $[\hat{a},\hat{a}^{\dagger}]=1$. Alternatively, one can write 
\begin{equation}
\hat{\Phi}=\Phi_{\mathrm{zpf}}\left(\hat{a}^{\dagger}+\hat{a}\right), \quad \hat{Q}=i Q_{\mathrm{zpf}}\left(\hat{a}^{\dagger}-\hat{a}\right) ,
\end{equation}
where $\Phi_{\mathrm{zpf}}=\sqrt{\hbar / 2 \omega C}=\sqrt{\hbar Z / 2}$  is the zero point spread of the phase variable, $\omega$ is the small oscillation frequency of the oscillator, and $\Phi_{\mathrm{zpf}}Q_{\mathrm{zpf}}=\hbar/2$. Insisting on the parallel with the mechanical oscillator, $\Phi_{\mathrm{zpf}}$ is the electrical analog to the ground state position uncertainty and $Q_{\mathrm{zpf}}$ corresponds to the ground state momentum uncertainty. The capacitance $C$ then plays the role of the particle's mass. 

In the case of the SNAIL transmon used in Ref.~\cite{Frattini2022}, the Hamiltonian of the driven circuit, which is built by an arrangement of a few Josephson junctions, reads
\begin{equation}
\begin{aligned}
\label{eq:timedep}
\frac{\hat{H}(t) }{ \hbar}=\omega \hat{a}^{\dagger} \hat{a} &+\sum_{m=3}^{\infty} \frac{g_m}{m}\left(\hat{a}^{\dagger}+\hat{a}\right)^m \\
&-i \Omega_d\left(\hat{a}-\hat{a}^{\dagger}\right) \cos (\omega_d t).
\end{aligned}
\end{equation}
This is Eq.~(1) in Ref.~\cite{Frattini2022},  where the $g_n$'s are the circuit nonlinearities and the drive is defined by its amplitude $\Omega_d $ and its frequency $\omega_d$, which is fixed at two times the small oscillation frequency of the oscillator to create resonant squeezing. Since nonlinearity is sourced by  an arrangement of Josephson junctions in the SNAIL, the $g_n$ coefficients are of order $\Phi_{\mathrm{zpf}}^{n-2}$ \cite{Venkatraman2022}.  Additionally, the magnetic flux tuning of a SNAIL permits the tunability of the oscillator's nonlinearities \cite{Frattini2017}. In particular, one can tune the values of $g_3 (\Phi_B)$ and $g_4 (\Phi_B)$ rather accurately. For the sake of this discussion, we will approximate the impedance of the circuit as independent from the magnetic flux.

The static effective Hamiltonian describing the system in these conditions is given by 
\begin{equation}
\label{eq:HSK}
 \frac{\hat{H}}{\hbar} =  - K \hat{a}^{\dagger 2} \hat{a}^2 + \epsilon_2 (\hat{a}^{\dagger 2} +  \hat{a}^2),    
\end{equation}
where, from the microscopic theory introduced in \cite{Frattini2017}, we can write the Kerr constant as $K = -\frac{3g_4}{2} +  2\frac{10g_3^2}{3\omega_d}$ and $\epsilon_2 = g_3\frac{4\Omega_d}{3\omega_d}$ \cite{Frattini2022}.  This Hamiltonian is the quantum optical analog of a double-well potential \cite{Wielinga1993} and the number of levels inside the wells is given by $N=\xi/\pi$ \cite{Frattini2022}, where $\xi=\epsilon_2/K$ is the control parameter. 

The Hamiltonian in Eq.~(\ref{eq:HSK}) can be factorized to read 
$$
\hat{H} = -K(\hat{a}^{\dagger 2} - \epsilon_2/K) (\hat{a}^2 - \epsilon_2/K).
$$
This means that the coherent states 
$| \pm \alpha \rangle$,
with $(\pm \alpha)^2 = \sqrt{\epsilon_2/K}$, are both degenerate with eigenenergy zero. Since $-\hat{H}$ is positive semidefinite, then $| + \alpha \rangle $ and $| - \alpha \rangle $ are degenerate ground states of the system and can be thought of as the ground states of the double well. Note that the bonding and antibonding superpositions
$\propto(|+\alpha \rangle \pm |-\alpha \rangle)$
are exactly degenerate for all well-depths. This implies that there is no tunnel splitting between the well ground-states. This is a peculiarity of our Hamiltonian that has important consequences for the dynamics~\cite{VenkatramanARXIV,PradoARXIV}.
Beyond the ground state, it is only in the classical limit $\xi \rightarrow \infty$, that the degeneracy is total for the excited states. For finite values of $\xi$, the excited state splitting is reduced exponentially with $\xi$.

We note that flux tuning a SNAIL circuit allows for a Kerr-free point \cite{Sivak2019}, where the Kerr constant is null. We express this condition by writing $K=\Phi_{\mathrm{zpf}}^{2}\kappa(\Phi_B)$, with $\kappa(\Phi_B)$ a function crossing zero. In addition, the third-order nonlinearity responsible for the generation of squeezing remains essentially constant in the vicinity of the Kerr-free point, so we can write the scaling of the control parameter $\xi$ as a function of the experimentally controllable variables, 
\begin{equation}
\xi = \frac{\epsilon_2}{K} \propto \frac{\Omega_d}{\Phi_{\mathrm{zpf}} \kappa(\Phi_B)}.
\end{equation}
With this expression, it is clear that the value of $\xi$ can be increased 
in three different ways. One can (i) reduce the impedance of the circuit, thus reducing $\Phi_{\mathrm{zpf}}$, (ii) increase the microwave power of the squeezing drive $\Omega_d$, or (iii) approach the Kerr-free point by in situ magnetic flux tuning $\Phi_B$.

\subsection{Classical limit}

As we wrote in Methods, the experimental system admits an approximate classical description if it is initialized in a coherent state and for as long as the Hamiltonian phase space surface produces only a linear force (a quadratic Hamiltonian) over the spread of the evolving state~\cite{Curtright2013}. This means that the dynamics will be generated by the Poisson bracket, since the Moyal corrections can be neglected, and no phase space interference effects will develop. This can be achieved by reducing the fluctuations of the coherent state (increasing its ``mass", $\Phi_{\mathrm{zpf}}\rightarrow0$), or by making a comparatively large double-well system ($\Omega_d/\kappa(\Phi_B)\rightarrow\infty$). Note that reducing $\Phi_{\mathrm{zpf}}$ comes at the price of increasing the spread in the momentum coordinate. In a Hamiltonian with quadratic kinetic energy, like Eq.~(\ref{eq:timedep}), this comes at a minimal cost, since no nonlinearity is experienced along the momentum (charge) axis, and the Moyal corrections remain small. Note, however, that in the presence of a Kerr nonlinearity, the Hamiltonian has a nonlinear dependence on the momentum coordinate and the classical correspondence needs to be treated carefully. This justifies taking the classical limit as a system of increasing size, $\xi\gg1$, as in the main text, which can be achieved independently of the value of zero point fluctuations. 

In the absence of dissipation, this classical Hamiltonian  approximation breaks at sufficiently long times for most initial conditions. In turn, small amounts of dissipation enforce the classical dynamics \cite{Frattini2022,Zurek2004,Habib1998}.
As in \cite{Zurek2004,Zurek1995}, the system discussed here is not chaotic, but since for a state initialized near the ESQPT, the evolution can be approximated by a quadratic Hamiltonian (squeezing, $\epsilon_2\gg K)$, the exponential instability is a property of both the quantum and the classical models. The evolution can be approximated as classical until the phase space distribution folds on the quartic energy wall and develops phase interferences, such as those seen in the last snapshop of the last row of the Fig.~2e in the main text \cite{Zurek2003}. This quantum-classical divergence will be regularized in a timescale set by dissipation. The possibility to experimentally explore the quantum-classical correspondence in the squeeze-driven Kerr oscillator will be communicated elsewhere.

\section{Clustering of Eigenvalues and Static Observable}
\label{SM_cluster}

\begin{figure}[h]
 \bf{
	\centering
		\includegraphics[width=0.6 \textwidth]{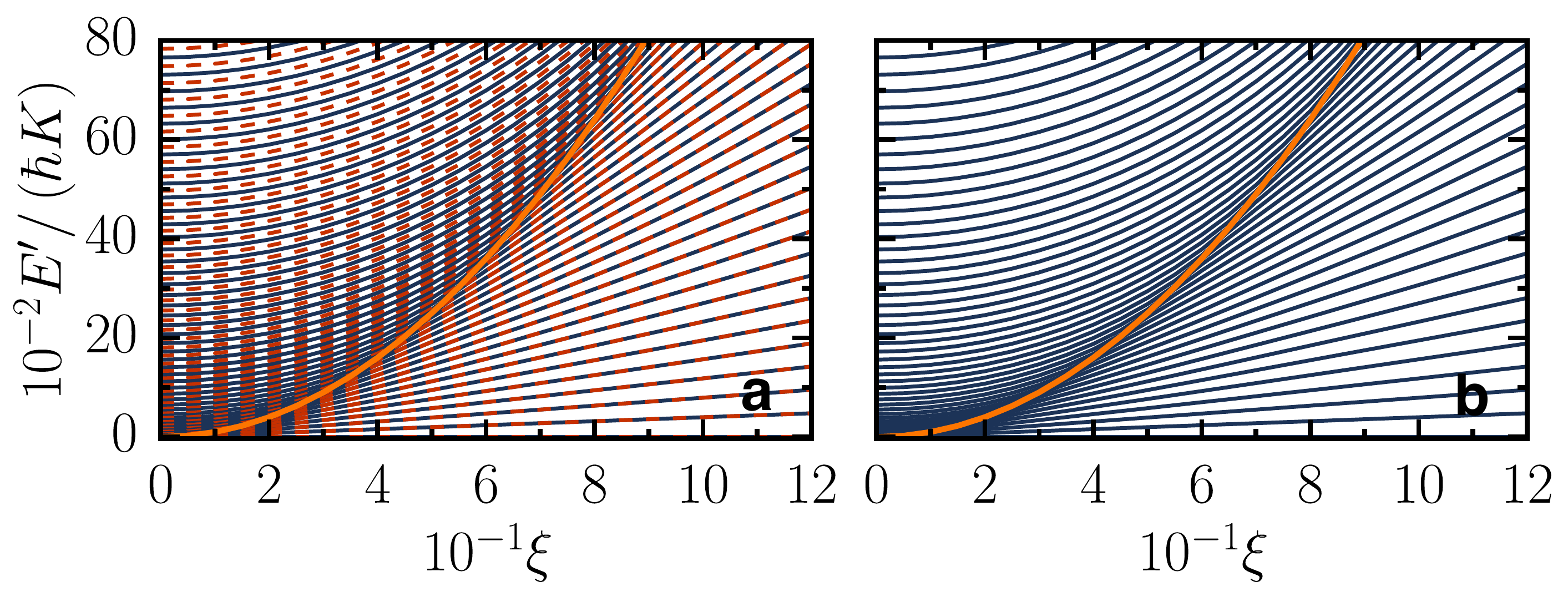}}
	\caption{Spectral kissing and eigenvalues clustering. {\bf a} {\normalfont Energy levels as a function of the control parameter $\xi$ for both parity sectors and {\bf b} for the even parity only.
	In {\bf a}: solid lines are for the energy levels in the even parity sector and dashed lines for the odd parity. The bright solid line in both panels marks the energy of the ESQPT, as given in Eq.~(3) of the main text.}
	}
		\label{fig01SM}
	\end{figure}

Supplementary Figure~\ref{fig01SM}a is identical to Fig.~1b in the main text. Supplementary Figure~\ref{fig01SM}b is the same as Supplementary Figure~\ref{fig01SM}a, but displayed for a single parity sector with the purpose of making it evident that the clustering of the eigenvalues  at $E'_{\text{ESQPT}}$ happens also in a single sector. The size ${\cal N}$ of the truncated Hilbert space here and everywhere in this work is chosen to guarantee the convergence of the energy levels analyzed.

\renewcommand{\figurename}{{\bf Supplementary Figure}}

\begin{figure}[h]
 \bf{
	\centering
		\includegraphics[width=0.6 \textwidth]{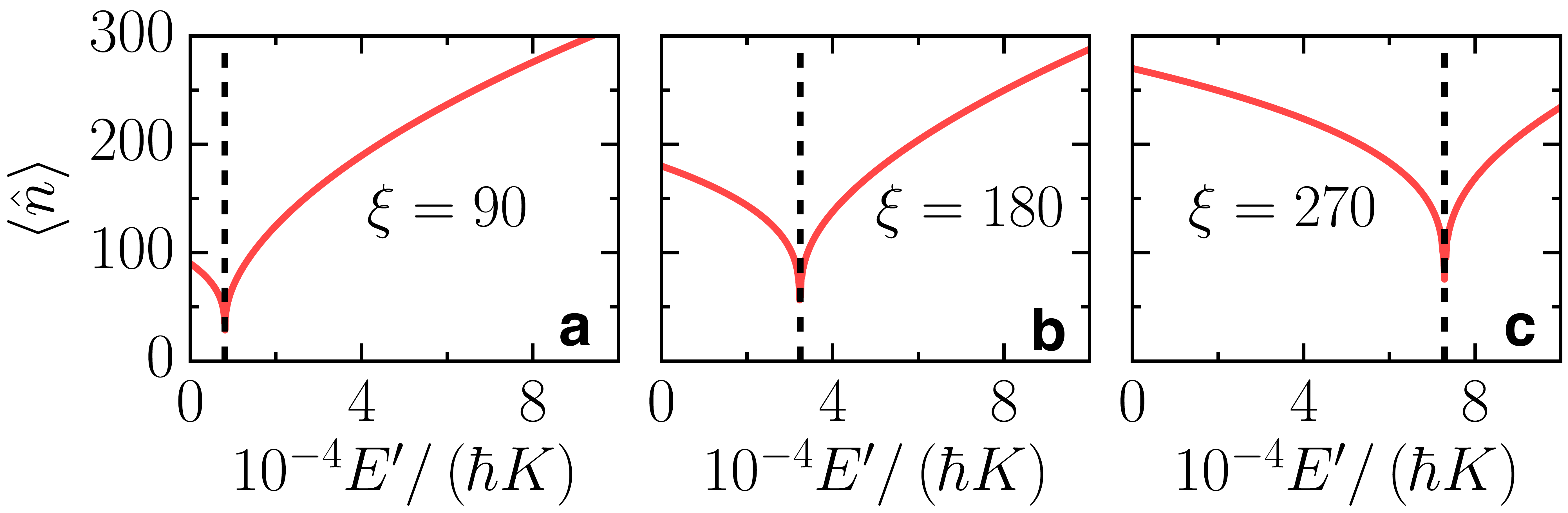}}
	\caption{Occupation number detects the ESQPT. {\normalfont Eigenstate expectation value of the occupation number as a function of the excitation energies for different values of the control parameter $\xi$, as indicated in the panels. An abrupt dip happens at $ E' \sim  E_{\text{ESQPT}}'$.}
	}
		\label{fig02SM_n}
	\end{figure}

Supplementary Figure~\ref{fig02SM_n} shows the eigenstate expectation value of the number operator, $\langle \hat{n} \rangle = \langle \psi |\hat{a}^{\dagger} \hat{a}|\psi \rangle$, as a function of the excitation energies. A dip is clearly seen at the ESQPT energy, $ E' \sim  E_{\text{ESQPT}}'$, which is caused by the fact that the eigenstate at this energy is localized in the Fock state $|0 \rangle$. The figure is analogous to that of the participation ratio in Figs.~1f-h of the main text.

\section{Density of States}
\label{SM_dos}

We can use the lowest-order term of the  Gutzwiller trace formula~\cite{GutzwillerBook}  to obtain a semiclassical approximation for the DOS,
\begin{eqnarray}
    \nu ({\cal E}) &=& \frac{1}{2\pi} \int dp dq \delta(H_{cl} - {\cal E} ),
    \label{Eq:DOS}
\end{eqnarray}
where $H_{cl}$ is given by
\begin{equation}
 H_{cl} =  -  \frac{K_{cl}}{4 } (q^2+p^2)^2 + K_{cl} \xi_{cl} ( q^2-p^2) .
 \label{EqHcl_SM}
\end{equation}
This is Eq.~(2) of the main text and, with the proper sign, it corresponds to  Eq.~(11) in Methods. To evaluate the previous integral, we use the general property of the Dirac delta,
\begin{equation}
\int_{\mathbb{R}^n}f(\textbf{x})\delta(g(\textbf{x}))d\textbf{x}=\int_{g^{-1}(0)}\frac{f(\textbf{x})}{|\nabla g|}d\sigma(\textbf{x}),
\end{equation}
where the integral on the right is over $g^{-1}(0)$ and the $(n-1)$-dimensional surface defined by $g(\textbf{x})=0$. Employing the  property of the Dirac delta in the Gutzwiller formula, we have
\begin{equation}
\nu ({\cal E})=\dfrac{1}{2\pi}\int_{q\in\Omega_{\cal E}}\dfrac{dq}{2\sqrt{(2\sqrt{K_{cl}\,u({\cal E})}-(\lambda+K_{cl}q^2))u({\cal E})}},
\end{equation}
where $u({\cal E})={\cal E}-{\cal E}_\text{min}+\lambda q^2$, $\lambda=2K_{cl} \xi_{cl} $, and $\Omega_{\cal E}$ is the set of values of $q$ for which there is at least one solution of the equation $H_{cl}(q,p)={\cal E}$.

\section{Lyapunov Exponent}
The linear analysis around the center and the hyperbolic points gives us information about the qualitative behavior close to those points. In particular, for the Hamiltonian in Eq.~(\ref{EqHcl_SM}), the linearized Hamilton equations around a critical (stationary) point  $\{q_c, p_c\}$ satisfy
\begin{eqnarray}
\begin{pmatrix}
 \dot{q} \\
\dot{p}
\end{pmatrix}
=   \begin{pmatrix}
-2K_{cl} q_c\,p_c & \hspace{-0.9 cm } -2K_{cl}\xi_{cl} \!-\! K_{cl}(q^2_c+3p^2_c) \\
-2K_{cl}\xi_{cl} \!+\! K_{cl}(3q^2_c+p^2_c) & \hspace{-0.9 cm }  2K_{cl} q_c\,p_c \\
\end{pmatrix}
\hspace{-0.1 cm }
\begin{pmatrix}
q\!-\!q_c \\
p\!-\!q_c
\end{pmatrix}. 
\nonumber
\end{eqnarray}
In the equation above, $cl$ stands for ``classical'' and $c$ for ``critical''.
The stability or instability around  $\{q_c, p_c\}$ is given by the eigenvalues $a_i$ of the matrix constructed by the linear system. If the eigenvalues of the matrix are real, then the Lyapunov exponent is equal to $\text{max}(a_i)$. 

For the specific case of the hyperbolic point $\{q_c, p_c\} = \{0, 0\}$,  the linear system is given by
\begin{equation}
\begin{pmatrix}
\dot{q}  \\
\dot{p}
\end{pmatrix}=
\begin{pmatrix}
0 & -2K_{cl}\xi_{cl} \\
-2K_{cl}\xi_{cl} & 0 \\
\end{pmatrix}
\begin{pmatrix}
q\\
p
\end{pmatrix}
\label{EqS:saddle}
\end{equation}
and the Lyapunov exponent is 
\begin{equation}
\lambda=2K_{cl}\xi_{cl}.
\end{equation}
At the vicinity of the hyperbolic point, the dynamics is dominated by the squeezing part of the Hamiltonian, $H_{cl} \approx K_{cl}\xi_{cl}  (\hat{q}^2 - \hat{p}^2)$, and the solution of Eq.~(\ref{EqS:saddle}) gives
\begin{equation}
\begin{pmatrix}
q\\
p
\end{pmatrix}=c_1
\begin{pmatrix}
1\\
1
\end{pmatrix}\text{e}^{\lambda t}
+c_2
\begin{pmatrix}
-1\\
1
\end{pmatrix}\text{e}^{-\lambda t} ,
\end{equation}
where $c_1$ and $c_2$ are constants. Two directions of evolution control the dynamics in the phase space, the direction $\begin{pmatrix}
1\\
1
\end{pmatrix}$ is affected by the positive exponential, as a repellor, and $\begin{pmatrix}
-1\\
1
\end{pmatrix}$ is affected by the negative exponential, as an attractor. The hyperbolic (saddle) point at the origin of phase space  is both a repellor and an attractor. A trajectory in the vicinity of this point moves towards and away from it. The quantum counterpart of this behavior is observed with the evolution of the Husimi function for the initial coherent state centered at O=\{0,0\}, which spreads rapidly at short times, but eventually folds back towards the initial distribution. This behavior is shown in Fig.~2e of the main text and is made yet more evident with the additional snapshots presented in the Supplementary Figure~\ref{figSM_ABC}.

\section{Quantum Dynamics}

The 6 initial coherent states that we consider are those listed in Eq.~(14) of the Methods in the main text.

\begin{figure*}[t]
\bf{
\centering
States A, B, and C \hspace{5.7 cm } State O
\vskip 0.1 cm
    \includegraphics[width=0.46\textwidth]{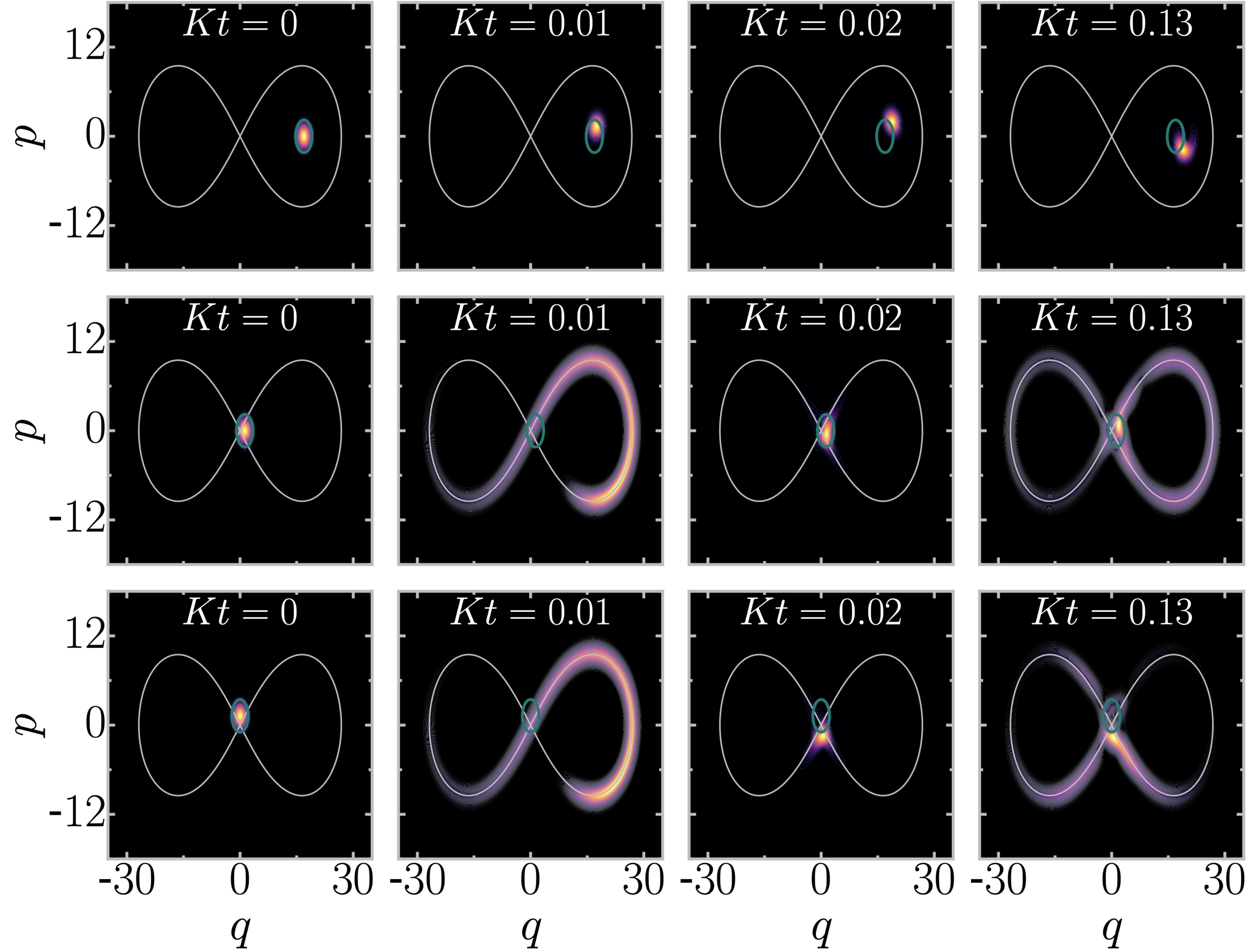} 
    \includegraphics[width=0.46\textwidth]{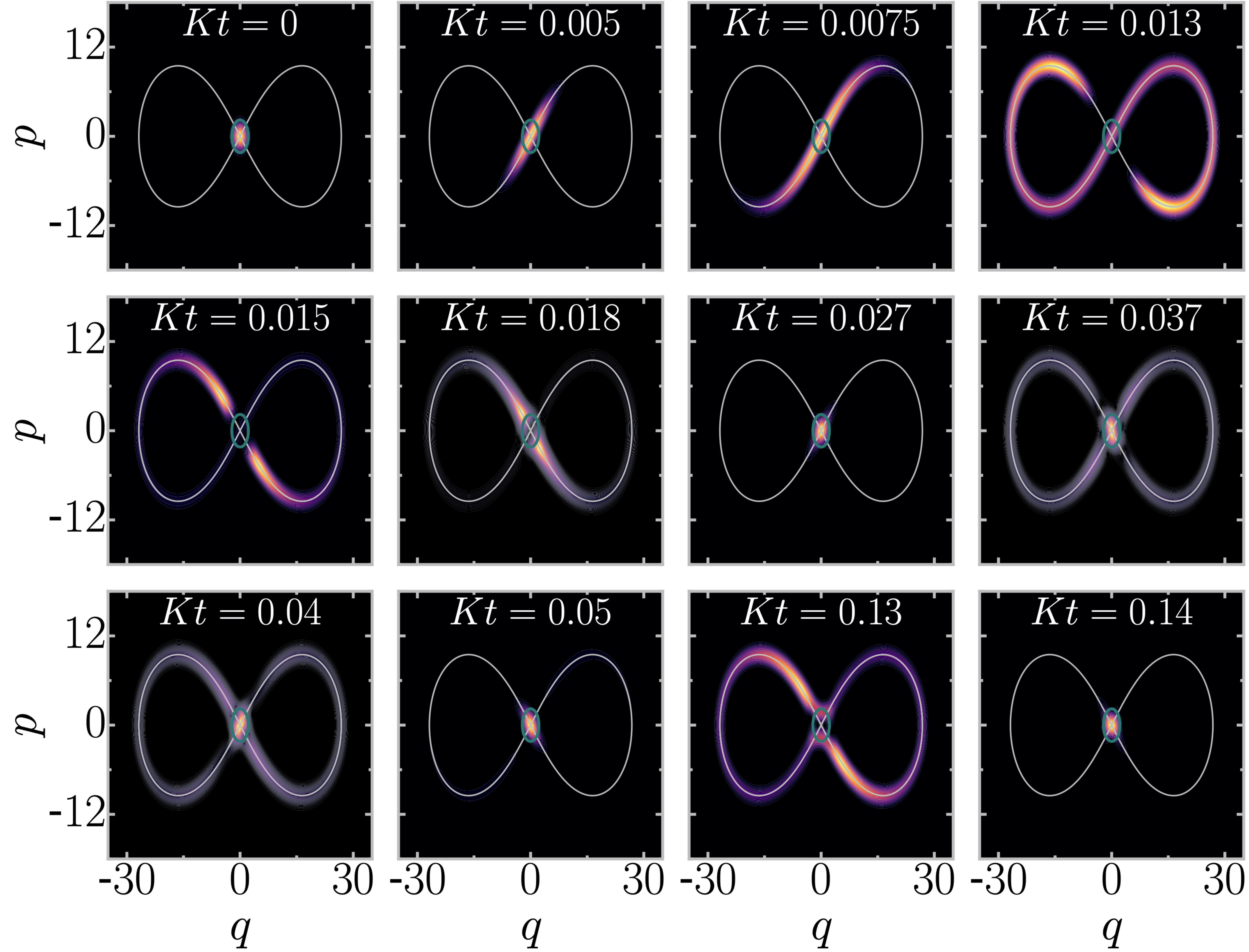} \\ 
    \vskip 0.1 cm
    \hspace{0.7 cm } State D \hspace{6.8 cm } State E
\vskip 0.1 cm
    \includegraphics[width=0.46\textwidth]{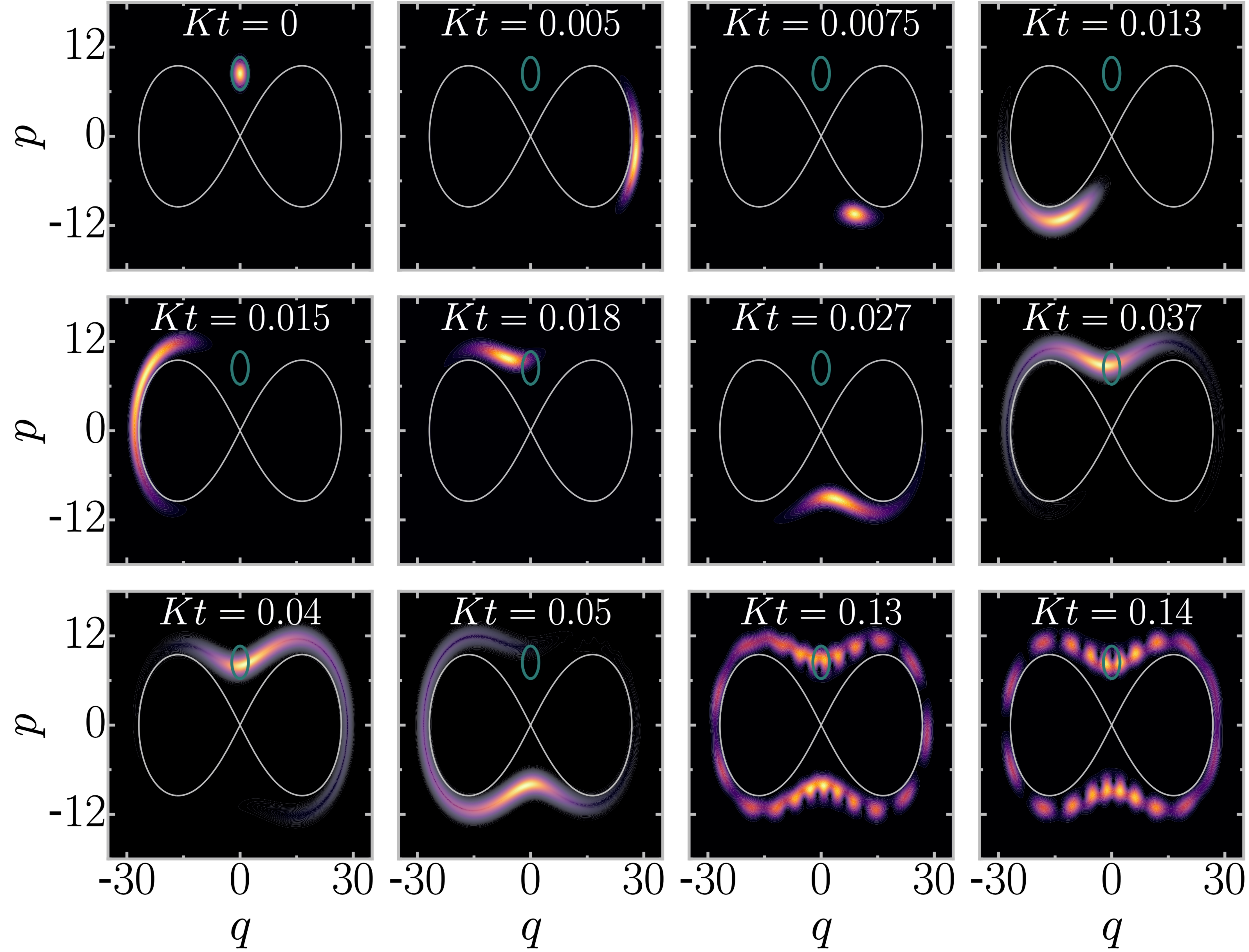} 
    \includegraphics[width=0.46\textwidth]{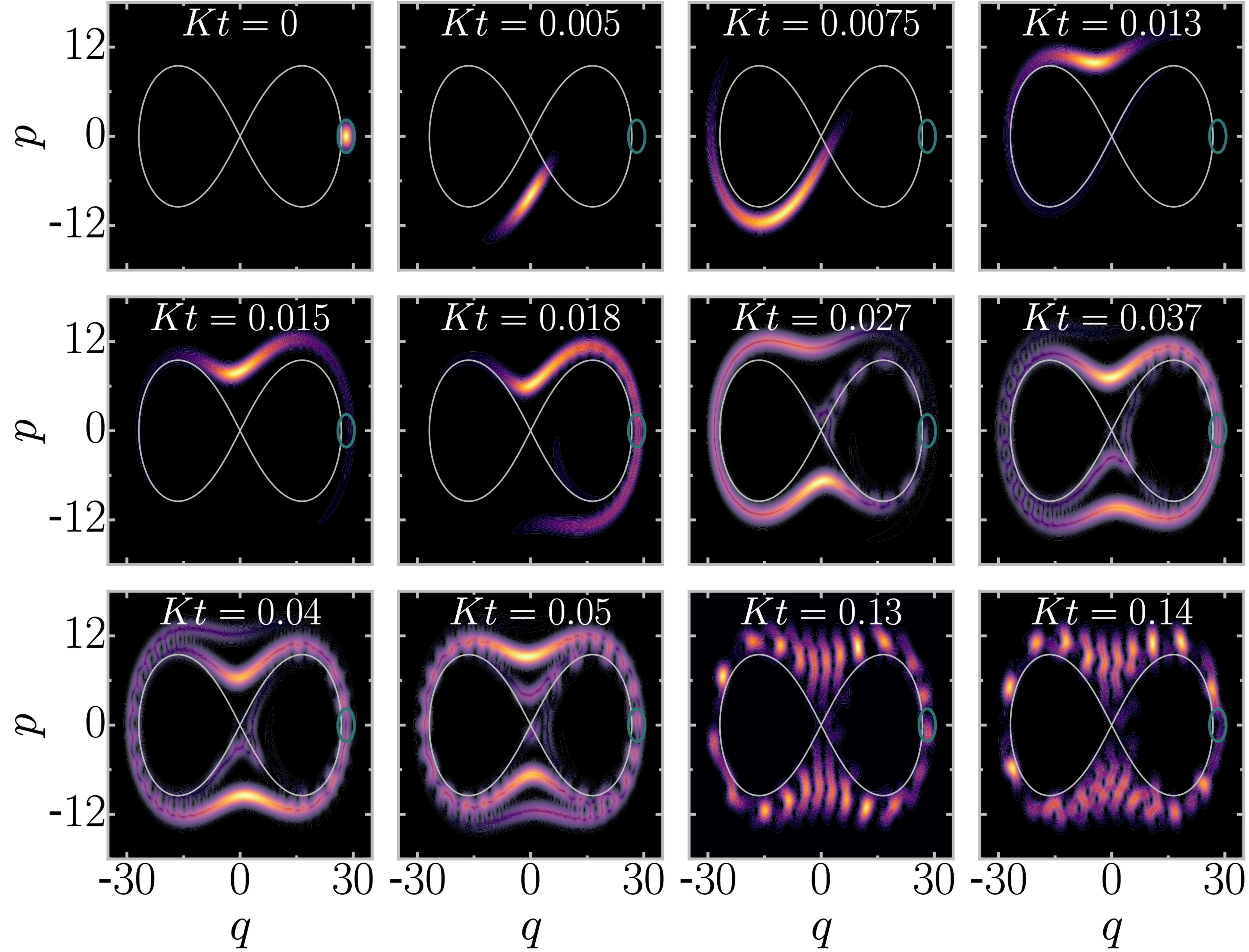}}
	\caption{Spread in phase space captured by the evolution of Husimi functions. {\normalfont Snapshots of the Husimi functions for the 6 initial coherent states investigated, as indicated in the titles; $\xi=\xi_{cl}=180$. On the top left, the snapshots in the first row of panels are for state A, in the second row they are for B, and the third row for C.}
	}
		\label{figSM_ABC}
	\end{figure*}

\subsection{Integral of the square of the Husimi function}

One can quantify how an initial coherent state spreads in the phase space by computing the integral of the square of the Husimi function,
\begin{equation}
    M_2^{\Psi(t)} = \frac{1}{2\pi} \int dq\, dp\, [Q^{\Psi(t)} (q,p)]^2, 
\end{equation}
where $N_{\text{eff}}=1$ and
\begin{equation}
Q^{\Psi(t)}(q,p)  \!=\! \frac{1}{2 \pi } \left| \sum_{n=0}^{\cal N} C_{n} (t) \text{e}^{-  \frac{(q^2+p^2)}{4}  }\frac{(q-ip)^n }{\sqrt{2^n  n!}} \right|^2 .
\label{EqSM_Q}
\end{equation} 

By writing the evolved state in the Fock basis, $|\Psi(t)\rangle = \sum_n C_n (t) |n \rangle$, one can solve the integrals exactly and obtain
\begin{align}
  M_2^{\Psi(t)} =& \frac{1}{\pi}\sum_{n_1, n_2, m_1, m_2} \frac{C_{n_1}(t) C^*_{n_2} (t) C_{m_1} (t) C^*_{m_2} (t)}{\sqrt{n_1!n_2!m_1!m_2!}}  \int d^2\alpha\, \text{e}^{-2|\alpha|^2}{\alpha^*}^{n_1+m_1}{\alpha}^{n_2+m_2} \nonumber \\
  =& \sum_{n_1, n_2, m_1, m_2} \frac{C_{n_1}(t) C^*_{n_2} (t) C_{m_1} (t) C^*_{m_2} (t)}{\sqrt{n_1!n_2!m_1!m_2!}} 
  \frac{(n_1+m_1)!}{2^{n_1+m_1+1}} \delta_{n_1+m_1,n_2+m_2}~. 
  \label{Eq:M2exact}
\end{align}

\subsection{Snapshots of the evolution of the Husimi functions}

Here, we present various snapshots of the evolution of the Husimi functions for the 6 initial coherent states investigated. The main features are summarized below.

The three rows of panels on the top left of Supplementary Figure~\ref{figSM_ABC} present snapshots of the Husimi functions for four instants of time for the initial coherent states  $|\Psi_A(0) \rangle$ (first row), $|\Psi_B(0) \rangle$ (second row), and $|\Psi_C(0) \rangle$ (third row). State A has very low energy and thus exhibits a very limited spreading around its initial region in the phase space. At $Kt=0.02, 0.13$, the distribution gets mostly out of the green ellipse that determines the initial state, so the value of $S_p^{(A)}(t)$ should become very small.

In contrast to $|\Psi_A(t) \rangle$, the Husimi distributions for $|\Psi_B(t) \rangle$ and $|\Psi_C(t) \rangle$ get squeezed, but do not fully leave the green ellipse. These two states present evolutions similar to the coherent state $|\Psi_O(t) \rangle$, since the two also start close to the hyperbolic point at the origin of the phase space. As mentioned in the main text, the fact that $|\Psi_B(t) \rangle$ evolves towards the region with negative values of $q$ is a quantum effect. The classical point B has a positive value of $q$ and is inside the separatrix, so classically, its orbit never reaches values of $q<0$.

The various snapshots of the Husimi functions for $|\Psi_O(t) \rangle$ (top right), $|\Psi_D(t) \rangle$ (bottom left), and $|\Psi_E(t) \rangle$ (bottom right) complement those displayed in the main text. The panels for $|\Psi_O(t) \rangle$ make evident the fast spread of this state, and also the subsequent alternating spread and contraction of its Husimi function. 

State E also spreads fast, because it is placed on the separatrix, although far from the origin. Just as for B and C, its exponential behavior is a  quantum effect.
Part of the quantum evolution of the coherent state E happens inside the separatrix and part of it is outside, creating two spreading fronts, as visible from the snapshots at $Kt=0.027, 0.037, 0.04$, and $0.05$. These different paths generate a complicated pattern of interferences, as shown for $Kt=0.13$ and $0.14$. 

Quantum interferences also appear for the initial coherent state $|\Psi_D(0) \rangle$. This state has a high energy that is equal to that of state E, but since $|\Psi_D(0) \rangle$ starts far from the separatrix, it does not spread as fast as $|\Psi_E(0) \rangle$; compare their Husimi functions, for example, at $Kt=0.037, 0.04, 0.05$.

\section{Quadratic behavior in time}

At very short times, the survival probability, FOTOC, and $M_2(t)$ present a quadratic behavior in time. The time interval for this behavior is derived by doing a Taylor expansion of the propagator $U(t)=\text{e}^{-i\hat{H}t}$, as discussed next.

\subsection{Survival probability}

At short times, the survival probability, can be written as~\cite{Tavora2017}

\begin{eqnarray}
    S_p(t) &=& \left| \langle \Psi(0)|\text{e}^{-i\hat{H}t}|\Psi(0) \rangle \right|^2 \nonumber \\ 
    & \approx & \left| \left\langle \Psi(0) \left|1 -i\hat{H}t - \frac{\hat{H}^2 t^2}{2}\right| \Psi(0) \right\rangle \right|^2 \nonumber \\
   &=& 1 - t^2 \left[ \langle \Psi(0)|\hat{H}^2|\Psi(0) \rangle - \langle \Psi(0)|\hat{H}|\Psi(0) \rangle^2 \right] \nonumber \\
  &=&  1 - \Gamma^2 t^2, \nonumber
\end{eqnarray}
where $\Gamma^2$ is the variance of the energy distribution of the initial state written in the energy eigenbasis, that is
$$
    \Gamma^2 = \sum_{k}|C_k^{(0)}|^2 (E_k^2 - E_0)^2,
$$
where 
$\hat{H}|E_k\rangle = E_k |E_k\rangle$,
$E_0 = \langle \Psi(0)|\hat{H} |\Psi(0)\rangle$,
and
$$C_k^{(0)} = \langle E_k |\Psi(0) \rangle.$$

Using the Fock basis $|n\rangle$ to write $\Gamma^2_O$ for the initial coherent state O, we have that
\begin{eqnarray}
    \Gamma^2_O &=& \sum_n \langle 0|\hat{H}|n\rangle \langle n|\hat{H}|0 \rangle - \langle 0|\hat{H}|0 \rangle^2 \nonumber \\
    &=&  \sum_{n \neq 0} \left| \langle n|\hat{H}|0 \rangle \right|^2, \nonumber
\end{eqnarray}
therefore,
\begin{eqnarray}
    S_p^{O}(t) &\approx& 1 - 2 \xi^2 K^2 t^2.
\end{eqnarray}
This implies that the survival probability for the state O decays quadratically for
\begin{equation}
    Kt < \frac{1}{\sqrt{2} \xi}.
\end{equation}

	\begin{figure}[h]
 \bf{
	\centering
		\includegraphics[width=0.6\textwidth]{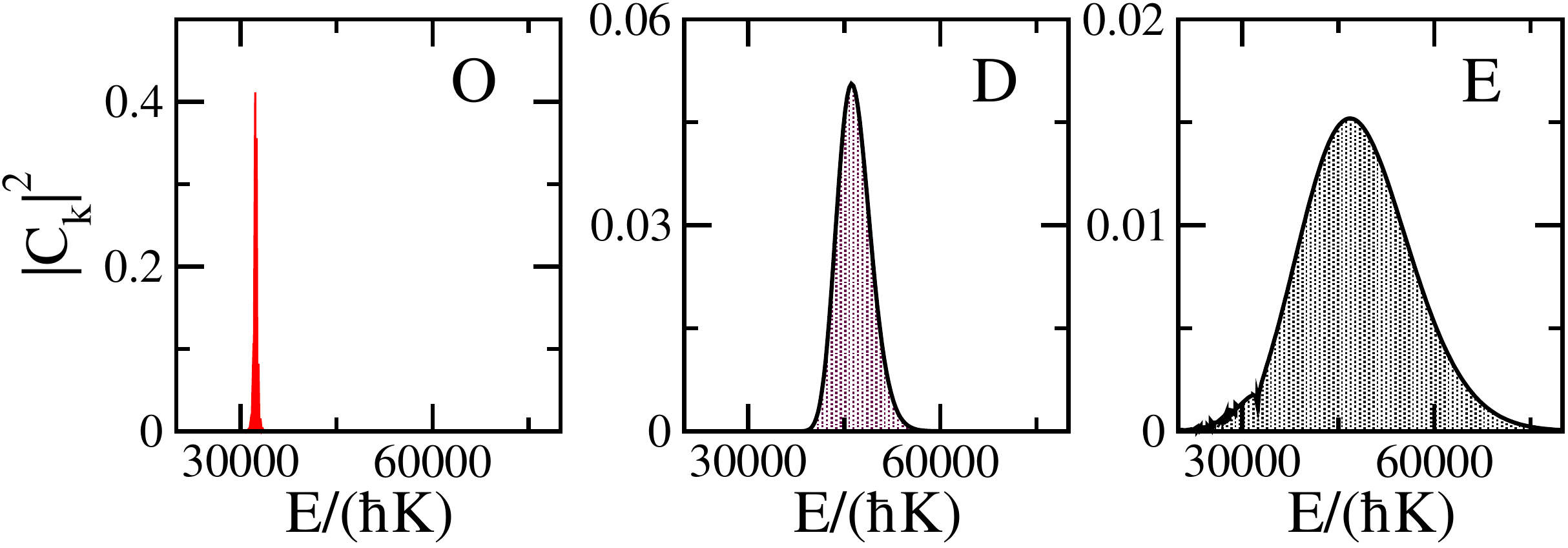}}
	\caption{Energy distribution of initial coherent states. {\normalfont The coherent states are centered at point O, D, and E, as indicated in the panels; $\xi =180$, $\hbar=1$.}
	}
		\label{FigSM_LDOS}
	\end{figure}

The derivation of $\Gamma^2$ for the other coherent states is analogous. As evident from the slowest decay of the survival probability for the initial coherent state $|\Psi_O(0) \rangle$ in Fig.~2d of the main text, this state has the smallest variance $\Gamma^2_O$. This happens because the corresponding classical point O is a stationary point.
The gradient and Laplacian of the Hamitonian vanish at O, so the initial diffusion constant for the Glauber coherent state $|\Psi_O(0) \rangle$ is the smallest one.

In the Supplementary Figure~\ref{FigSM_LDOS}, we show the energy distributions of the coherent states $|\Psi_O(0) \rangle$, $|\Psi_D(0) \rangle$, and $|\Psi_E(0) \rangle$. The width of the distribution for $|\Psi_O(0) \rangle$ is significantly narrower than for the other two states, as anticipated in the paragraph above.

Another feature observed in the Supplementary Figure~\ref{FigSM_LDOS} is the difference in the widths of the energy distributions for coherent states $|\Psi_D(0) \rangle$, and $|\Psi_E(0) \rangle$. Even though both initial states have the same energy, coherent state $|\Psi_E(0)\rangle$ is more spread out than $|\Psi_D(0)\rangle$, which explains why the survival probability $S_p^{E}(t)$ decays faster than $S_p^{D}(t)$, as seen in the Fig.~2d of the main text.

\subsection{FOTOC}

The same expansion of the propagator $U(t) = \text{e}^{-i \hat{H} t}$ can be extended to the analysis of the short-time evolution of the FOTOC, where one now needs to compute
$$
    \left\langle \Psi(0) \left| \left[1 +i\hat{H}t - \frac{\hat{H}^2 t^2}{2} \right] \hat{W} \left[1 -i\hat{H}t - \frac{\hat{H}^2 t^2}{2} \right] \right| \Psi(0) \right\rangle
$$
up to ${\cal O}(t^2)$
for $\hat{W} = \hat{p}$, $\hat{W} = \hat{p}^2$, $\hat{W} = \hat{q}$, and $\hat{W} = \hat{q}^2$. 

For the coherent state O, we find that
\begin{equation}
    F_{\text{otoc}}^{(O)} \approx 1 + 8 \xi^2 K^2 t^2,
\end{equation}
so its quadratic behavior holds for 
\begin{equation}
    Kt < \frac{1}{\sqrt{8} \xi}.
\end{equation}

\subsection{Short-time behavior of $\mathbf{M_2(t)}$}

To determine the duration of the quadratic behavior of $M_2(t)$, one needs to do the Taylor expansion for each component $C_n(t)$ in Eq.~(\ref{Eq:M2exact}), which becomes a tedious exercise even for the coherent state O. This timescale should again be dependent on the value of the control parameter, and we verify numerically that 
\begin{equation}
    Kt<\frac{1}{\xi}.
\end{equation}
is an upper bound.

\section{Exponential growth and infinite-time average of the FOTOC}

The exponential growth of the FOTOC for the coherent state $|\Psi_O(0) \rangle$ holds up to the Ehrenfest time ${\cal T} $ \cite{Shepelyansky2020}, which in our case is given by 
\begin{equation}
K {\cal T} \sim - 0.0027 + \ln (\xi)/(2 \xi) .  
\label{Ehrenfest}
\end{equation}
In the Supplementary Figure~\ref{FigSMmax}a, we show numerical results for $K{\cal T}$  as a function of $\xi$, and we find very good agreement with the expression in Eq.~(\ref{Ehrenfest}). Numerically, the Ehrenfest time is estimated as the point where $F_{\text{otoc}}^{(O)}(t)$ first reaches its highest value.

	\begin{figure}[h]
 \bf{
	\centering
\includegraphics[width=0.7\textwidth]{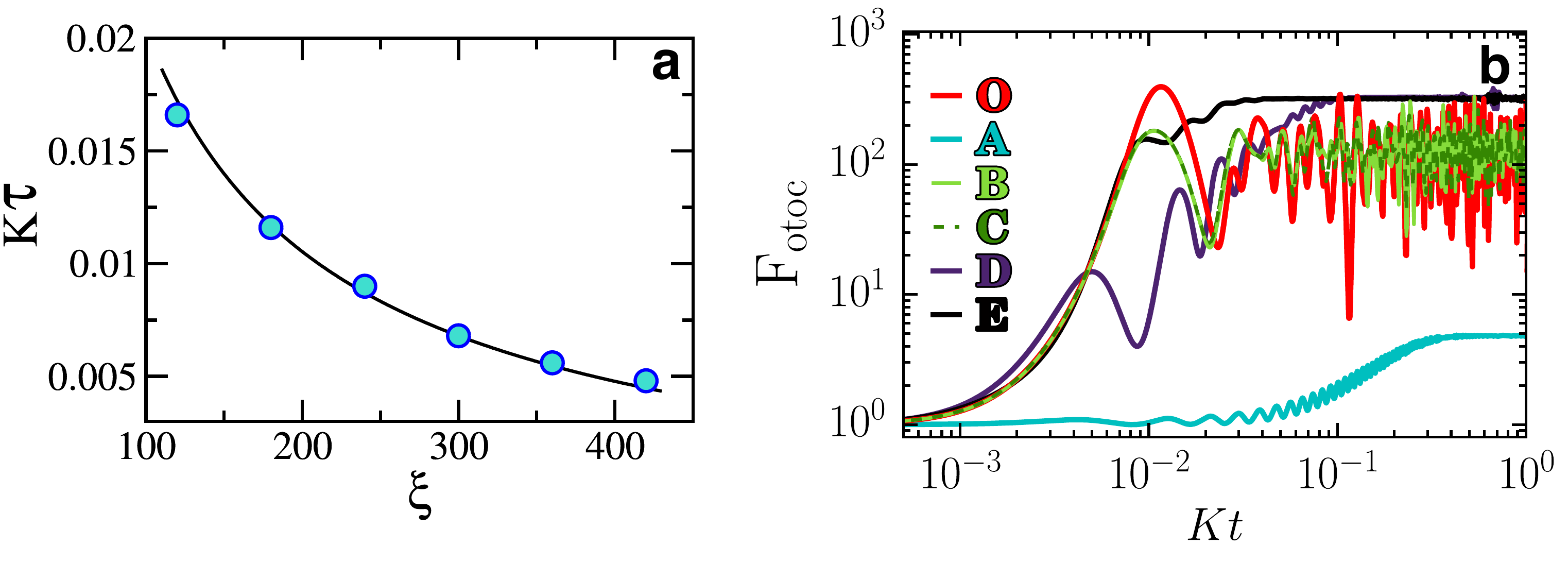}}
	\caption{Lyapunov time and long-time behavior of the FOTOC. {\normalfont {\bf a} Lyapunov time $ {\cal T}$ corresponding to the time when the FOTOC for the coherent state $|\Psi_O(0) \rangle$ first reaches its maximum value as a function of $\xi$. Symbols are for the numerical results and the solid line corresponds to the expression in  Eq.~(\ref{Ehrenfest}). {\bf b} Evolution of the FOTOC up to long times for the 6 initial coherent states considered in this work. The highest saturation value is reached by the two initial coherent states with the highest energies, $|\Psi_D(0) \rangle$ and $|\Psi_E(0) \rangle$.}
	}
	\label{FigSMmax}
	\end{figure}

In the Supplementary Figure~\ref{FigSMmax}b, we show results for the FOTOC for the same states shown in Fig.~2b of the main text, but up to longer times. We observe that $F_{\mathrm{otoc}}^{(A)}(t)$ saturates at the smallest value, because $|\Psi_A(0)\rangle$ has the lowest energy. 
$F_{\mathrm{otoc}}^{(O)}(t)$, $F_{\mathrm{otoc}}^{(B)}(t)$, and $F_{\mathrm{otoc}}^{(C)}(t)$ saturate at an intermediate and very similar value, since the states $|\Psi_O(0)\rangle$, $|\Psi_B(0)\rangle$, and $|\Psi_C(0)\rangle$ have similar intermediate energies. Among these three states, $F_{\mathrm{otoc}}^{(O)}(t)$ fluctuates the most. The infinite-time averages for $F_{\mathrm{otoc}}^{(D)}(t)$ and $F_{\mathrm{otoc}}^{(E)}(t)$ are equal and the highest among the six states, because $|\Psi_D(0)\rangle$ and $|\Psi_E(0)\rangle$ have an equal energy that is higher than that of the other six states. The temporal fluctuations of $F_{\mathrm{otoc}}^{(D)}(t)$ and $F_{\mathrm{otoc}}^{(E)}(t)$ are also much smaller than those for $F_{\mathrm{otoc}}^{(O),(B),(C)}(t)$.  

The results in the Supplementary Figure~\ref{FigSMmax}b indicate that, despite exhibiting the longest exponential growth and  the highest degree of scrambling up to the Ehrenfest time, the initial coherent state $|\Psi_O(0)\rangle$ centered at the hyperbolic point does not maintain the largest degree of spreading at long times. After the Ehrenfest time, $F_{\mathrm{otoc}}^{(O)}(t)$ is surpassed not only by the FOTOC of the state $|\Psi_E(0)\rangle$, which has an overlap with the separatrix, but even by the FOTOC of $|\Psi_D(0)\rangle$, which is away from the separatrix, but has higher energy than $|\Psi_O(0)\rangle$. This raises the question of how to define the notion of ``scrambling'' and how it depends on the timescales. 

\def\bibsection{\section*{Supplementary References}}
\vskip 1 cm

%

\end{document}